\documentclass[iop]{emulateapj}

\newcommand{\zet}{$\zeta_{\rm CR}(N_{\rm H_2})$}
\newcommand{\Bism}{$B_{\rm ISM}$}
\newcommand{\Bdisk}{$B_{\rm disk}$}

\usepackage{amsmath}
\usepackage{natbib}
\usepackage{enumerate}
\shorttitle{EXCLUSION OF COSMIC RAYS IN PROTOPLANETARY DISKS}
\shortauthors{Cleeves, Adams and Bergin}

\begin{document}

\title{Exclusion of Cosmic Rays in Protoplanetary Disks:  Stellar and Magnetic Effects}

 \author{L. Ilsedore Cleeves\altaffilmark{1}, Fred C. Adams\altaffilmark{1,2}, and Edwin A. Bergin\altaffilmark{1}}

\altaffiltext{1}{Department of Astronomy, University of Michigan, 500 Church St, Ann Arbor, MI 48109}
\altaffiltext{2}{Department of Physics, University of Michigan, 450 Church St, Ann Arbor, MI 48109}
 
\begin{abstract}
Cosmic rays (CRs) are thought to provide an important source of ionization in the outermost and densest regions of protoplanetary disks; however, it is unknown to what degree they are physically present.  As is observed in the Solar System, stellar winds can inhibit the propagation of cosmic rays within the circumstellar environment and subsequently into the disk.  In this work, we explore the hitherto neglected effects of cosmic ray modulation by both stellar winds and magnetic field structures and study how these processes act to reduce disk ionization rates.  We construct a two-dimensional protoplanetary disk model of a T-Tauri star system, focusing on ionization from stellar and interstellar FUV,  stellar X-ray photons, and cosmic rays.  We show that stellar winds can power a Heliosphere-like analogue, i.e., a ``T-Tauriosphere,'' diminishing cosmic ray ionization rates by several orders of magnitude at low to moderate CR energies ($E_{\rm CR} \le 1$ GeV).  We explore models of both the observed solar wind cosmic ray modulation and a highly simplified estimate for ``elevated'' cosmic ray modulation as would be expected from a young T-Tauri star. In the former (solar analogue) case, we estimate the ionization rate from galactic cosmic rays to be $\zeta_{\rm CR} \sim (0.23 - 1.4) \times10^{-18}$ s$^{-1}$. This range of values, which we consider to be the maximum CR ionization rate for the disk, is more than an order of magnitude lower than what is generally assumed in current models for disk chemistry and physics. In the latter elevated case,
i.e., for a ``T-Tauriosphere,'' the ionization rate by cosmic rays is $\zeta_{\rm CR} \lesssim 10^{-20}$ s$^{-1}$, which is 1000 times smaller than the interstellar value. We discuss the implications of a diminished cosmic ray ionization rate on the gas physics by estimating the size of the resulting MRI dead zones. Indeed, if winds are as efficient at cosmic ray modulation as predicted here, short-lived radionuclides (now extinct) would have provided the major source of ionization ($\zeta_{\rm RN} \sim 7.3\times10^{-19}$ s$^{-1}$) in the planet-forming zone of the young Solar Nebula.
\end{abstract}


\section{Introduction} 
Ionization is one of the most fundamental processes that drives the physics and chemistry of young protoplanetary disks.  From the physical perspective, processes such as accretion and planet formation depend crucially on the ability of the disk to transport angular momentum. The primary mechanisms posited for transport are gravitational instability \citep{cameron1978,boss1997} and the magnetorotational instability \citep[MRI;][]{velikhov1959,chandrasekhar1960,balbus1991}.  For disks with masses similar to that of the minimum mass Solar Nebula \citep[$M_d \lesssim$ 0.05 $M_{\odot}$;][]{weidenschilling1977} gravitational instability is not expected to be efficient and thus MRI is thought to drive angular momentum transfer.  However, for MRI to operate, the predominantly neutral disk must be sufficiently coupled to the magnetic fields through frequent ion--neutral collisions.  Therefore ions are essential in setting the kinematic and turbulent properties of the disk, in turn impacting important physical processes such as accretion onto the star and planet formation. Indeed, it has been suggested that low-ionization MRI-inactive regions of the disk, a.k.a. ``dead zones'' \citep{gammie1996,matsumura2003}, provide a favorable safe haven for planetesimal formation from dust coagulation \citep{gressel2012}.

Futhermore, ionization plays an important role in the heating of circumstellar gas.  Models of ionization by cosmic rays and X-rays have shown that the primary and secondary electrons generated in these processes can be a significant source of heating through inelastic collisions with gas molecules \citep{glassgold2001,glassgold2004,nomura2007,glassgold2012}. This additional heating source can significantly raise gas temperatures in the innermost radii and the tenuous surface layers of the disk, thereby influencing the strength and opacity of the observed emission lines. 

Finally, ions are critical for powering gas-phase chemistry, which proceeds predominantly through ion-neutral reactions -- the main chemical pathways in the interstellar medium. At low temperatures ($T < $ 50 K), ion--neutral reactions typically have reaction rates orders of magnitude faster than neutral--neutral reactions \citep{watson1976}, and therefore are the dominant drivers towards gas-phase complexity and enhancing deuteration, e.g., though reactions with H$_{2}$D$^+$.

The circumstellar molecular reservoir of the disk will eventually provide the material that will feed young proto-Jupiters, thereby setting the initial chemical composition of the gas-giants.  The turbulent properties of the disk also affects the efficiency with which the disk forms rocky planets and gas-giant cores \citep[e.g.,][]{gressel2012}, impacting the formation of smaller, Earth-like planets.   As a result, ions, through dynamical, thermal, and chemical mechanisms, influence all aspects of planet formation. 

The primary sources of ionization present in the disk environment include stellar and interstellar UV radiation, stellar X-rays, decay of short-lived radionuclides, cosmic rays (CR), and thermal ionization. In the surface layers and close to the star ($R \lesssim 0.5$ AU), stellar UV and X-ray photons are the dominant ionizing agents. At large radii ($R \gtrsim 100$ AU) and in the dense disk midplane where the optical depth to stellar and interstellar radiation is extremely large, it is often assumed that cosmic rays permeate material with column densities $\Sigma_g \lesssim$ 100 g cm$^{-2}$ \citep{umebayashi1981}, providing a base level of ionization and permitting MRI driven turbulence in the outer disk \citep[e.g., ][]{gammie1996,pbc2011a}.  

The need for CR ionization presents an interesting question.  First, it is seen that within the Solar System the solar wind modulates CR protons with energies below $\sim$ 1 GeV \citep{gleeson1967,gleeson1968,webber1974} within a region called the Heliosphere \citep{davis1955,axford1963}. Results from the Voyager spacecraft show the Heliosphere extends out to distances of at least 121 AU \citep{krimigis2011,decker2012}, with the true boundary yet to be crossed.  Young stellar objects are intrinsically magnetically active with significant mass-loss rates, and therefore it would be unsurprising for a young star to produce an analogous region of decreased CR flux, i.e.,  a ``T--Tauriosphere.''   It is important to note that the background CR flux in massive star-forming regions can have an elevated CR flux, amplified by supernova interactions with nearby molecular clouds \citep{fatuzzo2006}.  

Second, the star formation process is seen to reshape the magnetic environment \citep[e.g.,][]{girart2006}, and as a result the presence of magnetic structure can also modify cosmic ray propagation \citep{padovani2011,rimmer2012}.  Therefore it is not clear that cosmic rays are indeed present at rates predicted for the diffuse ISM \citep[$\zeta_{\rm CR} \sim 10^{-16}$ s$^{-1}$;][]{mccall2003,indriolo2007,neufeld2010} or even at the levels predicted for dense molecular clouds \citep[e.g., $\zeta_{\rm CR} \sim 3 - 7 \times 10^{-17}$ s$^{-1}$;][]{black1990}.  It is the latter rate that is typically adopted in simulations of MRI turbulence and circumstellar chemistry.

For the first time, this work explores the potential for stellar winds and magnetic fields to exclude CRs from the protoplanetary disk environment. Although this effect is extraordinarily well-studied within our own Heliosphere, it is typically neglected in modeling of circumstellar disks. \citet{turner2009} were the first to attempt to account for this reduction by integrating the ISM cosmic ray spectrum above $E_{\rm CR} \ge$ 100 MeV.  As this current paper shows, however, the exclusion of cosmic rays should be significantly more efficient and will reduce cosmic ray fluxes at all energies $E_{\rm CR}$. Here we examine how both wind and magnetic processes modify the ionization rate in the deep planet-forming layers of the disk and show how this shielding, in turn, can affect the size of dead zones. This paper is only the first step toward a full understanding of these processes. In subsequent work (Cleeves, Bergin \& Adams, in preparation; Paper II), we plan to examine the effects of cosmic ray exclusion on molecular ion chemistry and make predictions for observable tests by the Atacama Large Millimeter/Submillimeter Array (ALMA).

The paper is laid out as follows.  In Section~\ref{sec:model}, we motivate the physical model and discuss the sources of ionizing radiation present in the disk.  In Sections~\ref{sec:winds} and \ref{sec:mag}, we discuss the impact of stellar winds and magnetic fields on the cosmic ray ionization rate, respectively. In Section~\ref{sec:deadz}, we examine the effects of low ionization rates in the context of dead zones. In Section~\ref{sec:morestuff}, we discuss the possibility of additional ionization from the decay of radionuclides.  Finally, in Section~\ref{sec:disc}, we summarize our results.

\section{The Model}
\label{sec:model}
\subsection{Physical Parameters}
\label{sec:diskmod}

\begin{figure*}
\begin{centering} 
\includegraphics[width=6.8in]{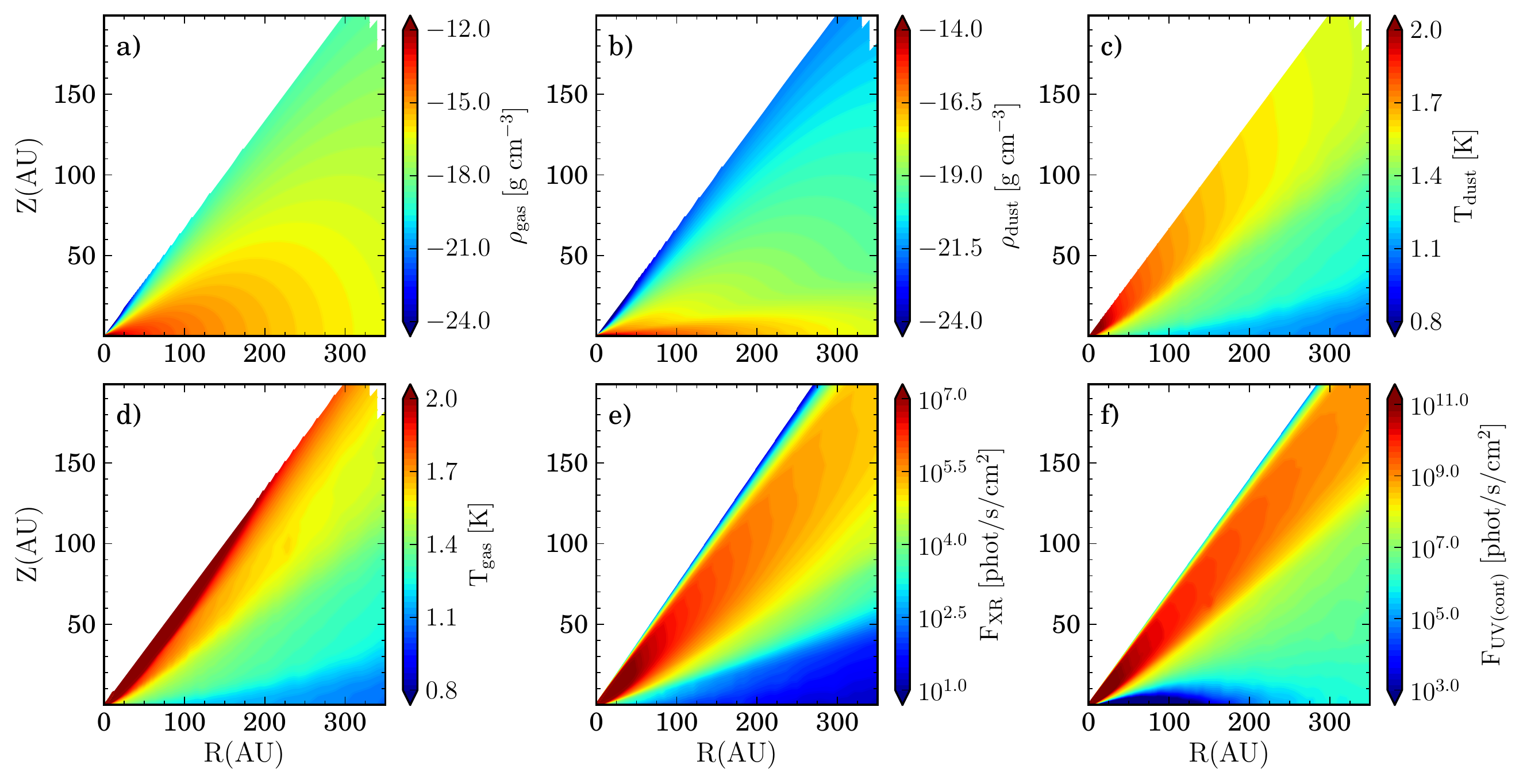}
\caption{Model a) gas density, b) dust density, c) dust temperature, d) gas temperature, and e) integrated X-ray- and f) UV-radiation fields. The UV flux is integrated between 930 -- 2000 \AA\ and has a spectral shape of TW Hydra \citep{herczeg2002,herczeg2004}.  The model X-ray flux has a total luminosity of $L_{\rm XR} = 10^{29.5}$ erg s$^{-1}$ between 0.1 -- 10 keV. }
\label{fig:diskmodel}
\end{centering}
\end{figure*}

Physical models of the dust and gas in protoplanetary disks have become increasingly complex \citep[e.g.,][]{dalessio1998,woitke2009,fogel2011}.  However, in this work we aim to isolate and demonstrate the effects of modifications to the cosmic ray flux and energy spectrum.  To do this, we have created a generic model of a T-Tauri disk and have altered the incident cosmic ray flux.  While the structure will depend in detail on the assumptions made for this incident flux (e.g., additional gas heating produced from secondary electrons), we have set out to understand the scope and intensity of cosmic rays throughout the disk in the context of ionization.  

With that established, we have created a 2D azimuthally symmetric model of a disk with the radiation transfer capabilities of the TORUS code \citep{harries2000,harries2004,kurosawa2004,pinte2009}. The 2D model structure and dust composition is fixed and then passively irradiated by the central star as described in detail below.  For our fixed density structure we have implemented a simple disk model of the form presented in \citet{andrews2011}:  
\begin{equation}
\Sigma_{g} (R) = \Sigma_{c} \left( \frac{R}{R_c}  \right) ^{-1} \exp{ \left[ -\left( \frac{R}{R_c} \right) \right]},
\end{equation}

\begin{equation}
h(R)=h_{c}{\Bigg(\frac{R}{R_{c}}\Bigg)}^{\psi}, \label{eq:hr}
\end{equation}

\begin{equation}
\rho_s (R,Z)=\frac{(1-f)\Sigma}{\sqrt{2 \pi}Rh} \exp{ \left[ - \frac{1}{2} \left(   \frac{Z}{h}  \right)^{2} \right]}, \label{eq:rs}
\end{equation} and
\begin{equation}
\rho_l (R,Z)=\frac{f\Sigma}{\sqrt{2 \pi}R \chi h} \exp{ \left[ - \frac{1}{2} \left(   \frac{Z}{\chi h}  \right)^{2} \right]}. \label{eq:rl}
\end{equation}

The first equation describes the radial surface density, a power-law with an exponential taper, where $\Sigma_c$ and $R_c$ are the characteristic surface density and radius of the profile, respectively. In this simple model, we mandate that both the gas and dust follow the same radial surface density profile, with the critical dust surface density equal to $\Sigma_d = \Sigma_g/100$ as per the standard ISM ratio.  The values for the parameters used are outlined in Table~\ref{tab:modpar} and are fixed unless otherwise noted.

Motivated by the {\it Spitzer} inference that small grains are not present in the upper layers of protoplanetary disks, i.e., are settled \citep{furlan2006}, we have incorporated the UV and X-ray optical effects of dust settling into our model.  Following the prescription described in \citet{andrews2011}, we have used an approximation to include vertical size segregation of dust grains by defining two distinct grain populations: one distribution corresponding to large, settled millimeter-sized grains and a second corresponding to a small micron-sized grain population, e.g., the ``atmosphere.''  Both the micron-sized grains and gas are distributed with a Gaussian profile of scale height $h(R)$ given by Equation~(\ref{eq:hr}) where $h_c$ is the characteristic scale height of the model and $\psi$ is the power law dependence of the scale height versus radial distance.  The large grains are distributed over a smaller scale height $\chi h$.  The value of $\chi < 1$ is fixed to 0.2, or, equivalently, large grains are distributed over 20\% of the scale height of small grains. While the vertically integrated gas mass to dust mass ratio is set to be 100, the gas to dust mass ratio is much larger in the surface layers and much smaller in the midplane. This geometry is expected due to grains preferentially settling out from the surface and increasing the mass of dust in the midplane, while simultaneously depleting the surface of dust.  

\begin{deluxetable}{lr}
\centering
\tablecolumns{2} 
\tablecaption{Stellar and disk model parameters.\label{tab:modpar}}
\tablewidth{0pt}
\tabletypesize{\footnotesize}
\startdata
 \hline
 \hline
 {\bf Stellar Model} \\ 
Stellar Mass & 1.06 $M_{\odot}$  \\
Stellar Radius & 1.83 $R_{\odot}$ \\ 
Stellar $T_{\rm eff}$ & 4300 K \\
$L_{\rm UV}$ &  2.9 $\times 10^{31}$ erg s$^{-1}$ \tablenotemark{a} \\
$L_{\rm XR}$  & 10$^{29.5}$ erg s$^{-1}$\\
 \\
{\bf Disk Model} \\ 
 $R_{\rm inner}$  &  0.1 AU \\
  $R_{\rm outer}$  &  400 AU \\
$M_{\rm dust}$  &    3.9 $\times 10^{-4}$ $M_{\odot}$ \\ 
$M_{\rm gas}$  &    0.039 $M_{\odot}$ \\ 
 a$_{\rm max}$ [atm.] & 1$\mu$m (15\%) \\
  & 10$\mu$m (85\%) \\
  a$_{\rm max}$ [midplane] & 1 mm \\
a$_{\rm min}$ & 0.005 $\mu$m \\
$f$ & 0.85 \\
$\chi $ & 0.2 \\
$\psi $ & 0.3 \\
$\Sigma_{c}$ & 3.1 g cm$^{-2}$ \\
$R_{c}$ & 135 AU \\
 $h_{c}$ & 12 AU
\enddata
\tablenotetext{a}{As computed from the observed FUV spectrum of TW Hya integrated between 930 and 2000 \AA\ \citep{herczeg2002,herczeg2004}.}
\end{deluxetable}

For the dust chemical composition, we implement a blend of 80\% astronomical silicates \citep{draine1984} and 20\% graphite; the mix is not varied between different size populations.  Both dust populations (atmosphere and midplane) are described by an MRN distribution \citep{mrn1977}, e.g., $n(a) \propto a^{-p}$ where $p = 3.5$, and have a characteristic minimum and maximum size $a_{\rm min}$ and $a_{\rm max}$.  The minimum size is furthermore fixed to be $a_{\rm min} = 0.005~\mu$m across all populations.  The small grains are broken up into two subsets, one with $a_{\rm max} = 1~\mu$m corresponding to 15\% small grains by mass and a second with $a_{\rm max} = 10~\mu$m accounting for 85\% of the mass in small grains.  The large grains are a single population with $a_{\rm max} = 1$ mm.  We have mandated that the mass in large grains, $f$, is 85\% of the total dust mass, see Equations~(\ref{eq:rs}) and (\ref{eq:rl}).

On this fixed model, TORUS solves for the dust thermal radiative equilibrium using the Lucy method (Lucy 1999).  For the central star, we adopt the following parameters: $M = 1.06 $ $M_{\odot}$, $R$ = 1.83 $R_{\odot}$, and $T_{\rm eff}$ = 4300 K.  Furthermore, the dust temperatures computed by TORUS are assumed to be equal to the gas temperatures.  We find that this assumption is acceptable as the bulk mass of the disk beyond $R = $ 0.5 AU is well below the temperature threshold for thermal ionization \citep[$T >$1000 K; ][]{fromang2002}.  To check this assumption, we have computed a simple equilibrium gas temperature by balancing heating from stellar X-rays against grain--gas collisional cooling \citep{glassgold2001,glassgold2004,qi2006} and find that the majority of the disk is below the thermal ionization threshold.  For completeness, we show the calculated gas temperature in Figure~\ref{fig:diskmodel}d., but emphasize that only the dust temperatures are used in the following calculations as this work focuses on ionization in the densest material where the dust and gas are thermally coupled.  The density and thermal structure of the disk model are plotted in Figure~\ref{fig:diskmodel}.

\subsection{Disk Ionization Processes}
\label{sec:ionprocess}

\subsubsection{Stellar Ionization}
\label{sec:ionmethod}

It is well known that young T-Tauri stars are bright X-ray and UV emitters \citep{feigelson1999}.  This emission irradiates the circumstellar environment and drives chemical, ionizing, and thermal physics.  The radiation from the star reaches the flared disk surface at a predominantly glancing angle and is scattered and absorbed by the circumstellar dust and gas. While the tenuous disk atmosphere is bathed completely by stellar UV and X-ray radiation, the large line-of-sight optical depths to the midplane hinder ionizing photons reaching the very densest outer regions (Figure~\ref{fig:diskmodel}).  It is thus important to model the propagation of radiation in the disk as accurately as possible to determine the volume of disk gas that is exposed to the stellar ionizing photons \citep[e.g.,][]{glassgold1997,igea1999}.  In this section we quantitatively show the relative contribution from each of the dominant sources of ionization for our prototypical disk model.

\begin{figure}
\begin{centering}
\includegraphics[width=3in]{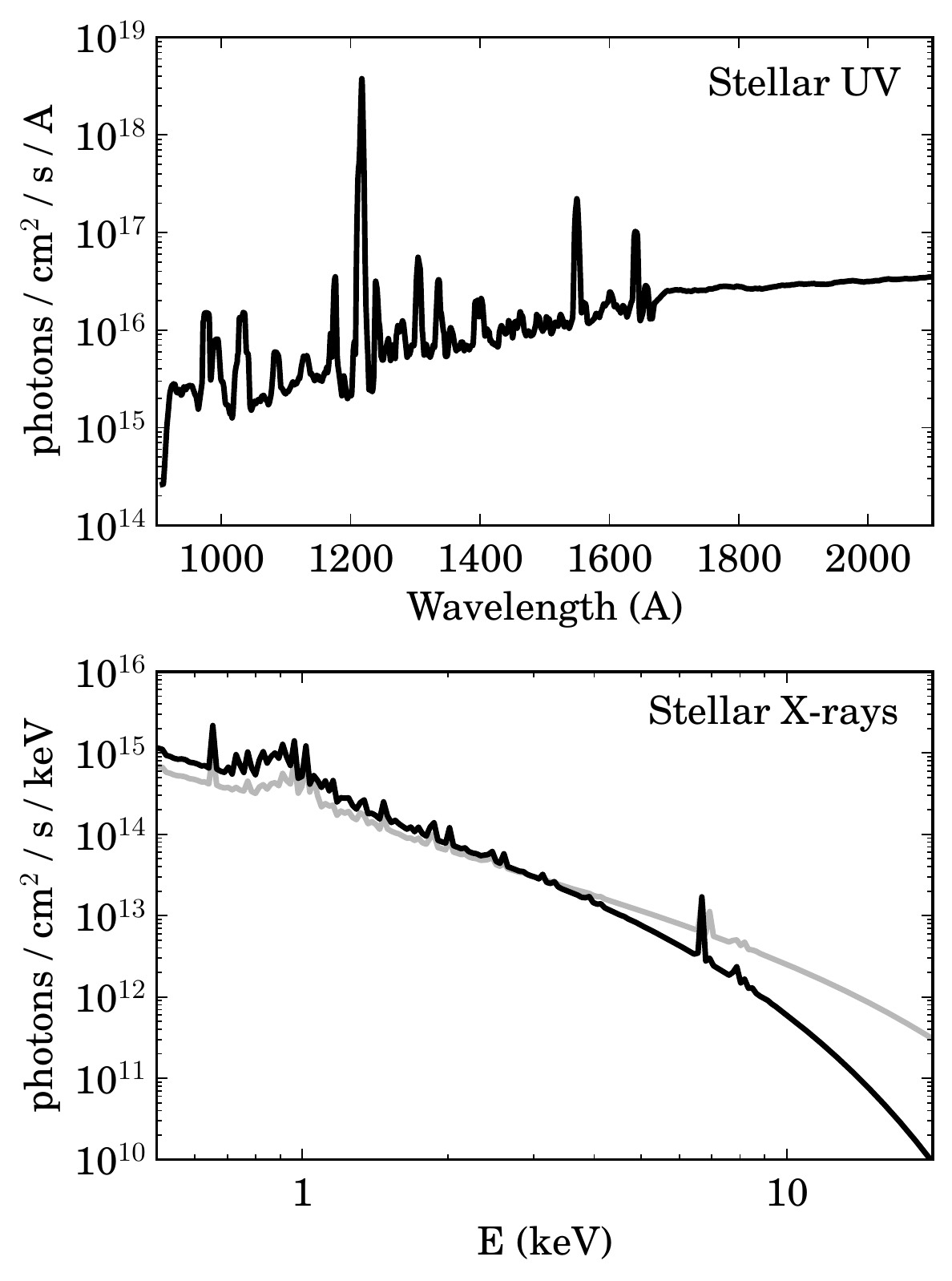}
\caption{Model UV and X-ray spectra from the central star taken at the stellar surface.  {\it Top:} Observed FUV spectrum of TW Hya \citep{herczeg2002,herczeg2004}; note the strong Ly-$\alpha$ emission at 1216 \AA.  {\it Bottom:} Model X-ray spectra as described in Section~\ref{sec:ionmethod}. The black line shows a ``characteristic'' T-Tauri X-ray spectrum and the grey line corresponds to a highly flaring T-Tauri star.  Both spectra are normalized to $L_{\rm XR}$ = 10$^{29.5}$ erg s$^{-1}$. We note that both models rise below $E \sim 1$ keV because we do not a include foreground absorption component in the model, typically required to reproduce observed foreground-extincted X-ray spectra. 
 \label{radfield}}
\end{centering}
\end{figure}

To determine the position dependent UV and X-ray fields, we employ a Monte Carlo radiative transfer code as described in \citet{bethell2011a, bethell2011b}. For the input stellar UV field, we adopt the observed spectrum of TW Hya \citep[Figure~\ref{radfield};][]{herczeg2002,herczeg2004}. The continuum opacity is set by the position dependent dust model and is computed via the Monte Carlo code at eight discrete wavelengths and interpolated between the range 930-2000 \AA.  In addition to these continuum points, we include the Monte Carlo radiative transfer of Lyman-$\alpha$ photons as described in \citet{bethell2011b}. In addition to dust scattering, Lyman-$\alpha$ photons undergo resonant isotropic scattering off hydrogen atoms, and therefore to determine the location of this hydrogen scattering layer the equilibrium H and H$_2$ abundances are computed using the method of \citet{spaans1997} and H$_2$ self-shielding functions of \citet{tielens1985}. 

The methodology of the X-ray radiative transfer is similar to that of the UV and incorporates both absorption and scattering of the X-ray photons.  We have used the updated X-ray cross sections of \citet{bethell2011b} that incorporate the recently updated solar abundances of \citet{asplund2009}. \citet{bethell2011b} found that even in the case of a fully dust-settled disk there exists an X-ray opacity ``floor'' from the gas.  Therefore, while settling does decrease the X-ray opacity, it is reduced by at most a factor of two at $\sim$ 1 keV.

The input X-ray spectra were generated using the MEKAL model \citep{liedahl1995} included in {\it XSPEC} \citep{arnaud1996} and correspond to a two-temperature optically thin thermal plasma.  In Figure~\ref{radfield}, the black line corresponds to a ``characteristic'' T-Tauri X-ray spectrum as measured at the stellar surface with temperature components corresponding to $T$ = 9 MK and 30 MK respectively \citep{getman2005,preibisch2005}.  On the same plot, the light grey line corresponds to the spectral shape typical of a T-Tauri in a X-ray high-flaring state \citep{getman2008}, normalized to the same luminosity as the quiescent spectrum with temperature components of $T$ = 12 MK and 100 MK respectively. 

The second spectrum is characteristic of T-Tauri stars with high X-ray luminosities ($L_{\rm XR} \gtrsim 10^{31}$ erg s$^{-1}$) and is thought to be the result of high stellar X-ray flare activity.  Observations of X-ray flaring sources show that during this period the X-ray spectrum furthermore becomes characteristically harder \citep{getman2008}.  In the following calculations we consider the ``characteristic'' (Fig.~\ref{radfield}, black line) X-ray spectrum in our benchmark model, but in Section~\ref{sec:extremeX} we consider the implications for an enhanced $L_{\rm XR}$ and harder X-ray spectrum.

The photon propagation is treated using the Monte Carlo code of \citet{bethell2011b}, where X-ray photons originate from the central star and then scatter/absorb as they propagate through the disk. The radiative transfer is computed at energies ranging $E$ = 1 -- 20 keV in 1 keV intervals. We note that some previous papers have adopted the assumption that X-rays are generated in accretion streams originating high above the stellar surface and therefore are able to ``shine down'' onto the disk. For example, \citet{igea1999} assume X-rays originate at a height of $z \approx 10$ $R_{\odot}$ above the midplane. However, even if this is the case, beyond 1 AU the difference in incidence angle between photons coming from the stellar surface and from an accretion stream would be $\lesssim 3^{\circ}$; thus, the approximation of a point source origin is satisfactory beyond 1 AU. The results of the UV and X-ray transfer are shown in Figure~\ref{fig:diskmodel} (e and f). 

\begin{figure*}
\begin{centering}  
\includegraphics[width=6.6in]{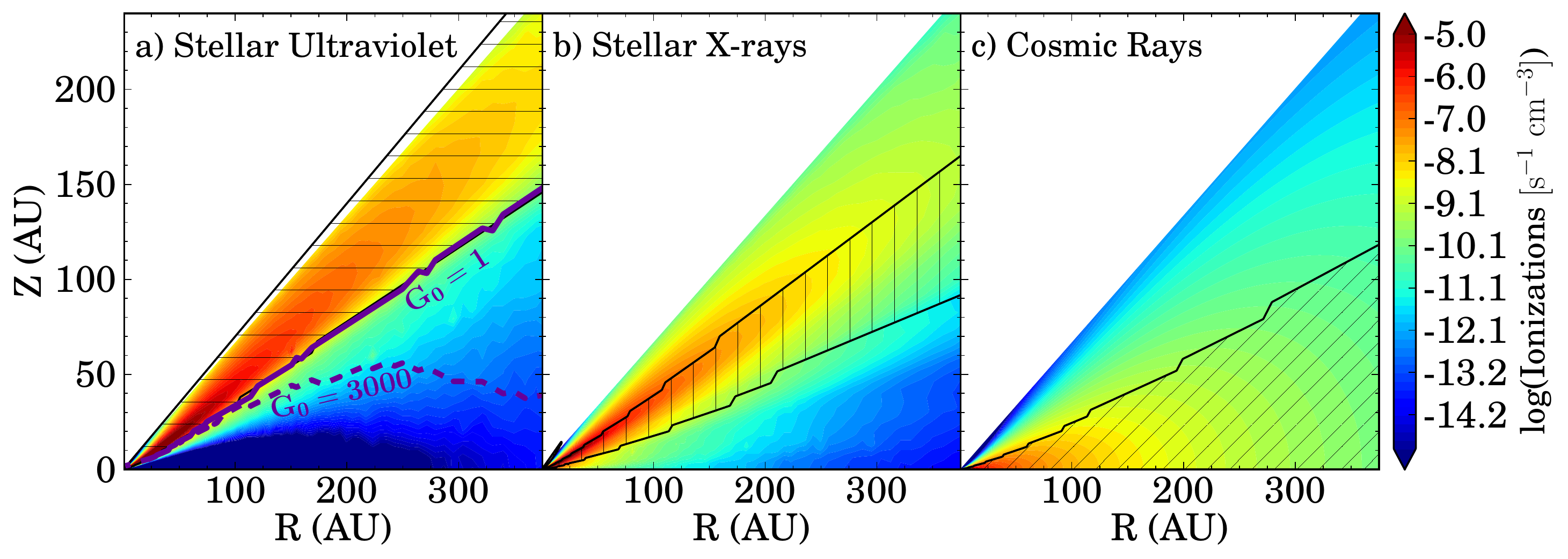}
\caption{Relative contribution of stellar UV, X-rays, and cosmic rays to the total ionization rate.  Colored contours show the volumetric ionization rate due to each source on the same scale. Hatched region delineates region of the disk where each respective source of ionization provides $> 30 \%$ of the total ionizations per unit time per unit volume.  The filled contours are a) FUV ionization of C, $n_{\rm C} \zeta_{\rm UV}$; b) X-ray ionization of H$_2$, $ n_{\rm H_2} \zeta_{\rm XR}$; c) cosmic ray ionization of H$_2$, $ n_{\rm H_2} \zeta_{\rm CR}$, for a standard ISM cosmic ray ionization rate $\zeta_{\rm CR} = 5 \times 10^{-17} \exp{[-\Sigma / (96 \rm~g~cm^{-2})]}$.    See Fig.~\ref{fig:isrf} for details of the $G_0$ contours (thick purple lines).}
\label{fig:ion3pan}
\end{centering}
\end{figure*}

Ultraviolet ionization (with rate $\zeta_{\rm UV}$) acts largely on carbon in the upper atmosphere of the disk.  As an upper limit we assume all carbon is in C or CO with $\chi_{\rm C} = 1.4 \times 10^{-4}$ and approximate the C--CO transition layer by balancing CO photodissociation with CO formation \citep{nelson1997}. The UV volumetric ionization rate ($\zeta_{\rm UV}$ multiplied by the number density of carbon atoms), in units of s$^{-1}$ cm$^{-3}$, is shown in Figure~\ref{fig:ion3pan} (a). The volumetric ionization rate is a rough proxy for the electron production rate per unit volume (as opposed to comparing the less intuitive ionization rate per C against X-rays or cosmic rays per H$_2$).

X-ray photons have a higher penetration column compared to UV and can ionize denser regions, acting primarily on H$_2$ and He. We note that while gas and metals in the dust provide an important source of X-ray photon extinction, the ionization of H$_2$ produces the bulk of the electron/ion abundance, along with He to a lesser extent.  In Figure~\ref{fig:ion3pan} (b) we plot the volumetric ionization rate due to X-ray ionization of H$_2$ using the X-ray ionization cross sections of \citet{igea1999} for the case of a settled (segregated) disk and assume an energy $\delta \epsilon$ = 37 eV is necessary to produce an ion pair \citep{shull1985, voit1991,igea1999}, as a result, a 1 keV X-ray photon produces $\sim$ 27 ion pairs.

Taking the TW Hydra FUV field,  $L_{\rm XR} =10^{29.5}$ erg s$^{-1}$, $\zeta_{\rm CR} = 5 \times 10^{-17}$ s$^{-1}$ $\rm H^{-1}$ and $\Sigma_{\rm CR} =$ 96 g/cm$^{2}$ (see Section~\ref{sec:crgen} regarding CR parameters), we plot the volumetric ionization rate from each major ionization source on the same scale (Figure~\ref{fig:ion3pan}). The surface is clearly dominated by UV ionization of carbon (left panel), while at a deeper intermediate layer X-rays dominate (center panel).  The black hatched regions indicate where each source provides at least 30\% of the total ionizations respectively.

\subsubsection{Interstellar Ultraviolet Ionization}
\label{sec:isrf}
In addition to ultraviolet irradiation from the central star, the interstellar radiation field (ISRF) provides an external source of UV ionization.  For an isolated disk or low mass star forming region the interstellar field provides an omnidirectional incident UV flux $\int F_\nu d\nu$ = 1.6 $\times$ 10$^{-3}$ erg cm$^{-2}$ s$^{-1}$ between 912 -- 2000 \AA, corresponding to $G_0 = 1$ \citep{habing1968}.  For comparison, the integrated flux from the star at a distance of 100 AU has typical values of $G_0 =$ 240 -- 1500 \citep{bergin2004}, dropping off as $\propto r^{-2}$.  However, for a disk in a stellar cluster $G_0$ can be much higher, with values ranging from 300 -- 30,000, with a typical value of 3000 \citep{fatuzzo2008}.  As a result, the interstellar field in a cluster can rival the stellar FUV radiation in the outermost regions of the disk.  This complication will certainly have important chemical implications but in this section we focus on the implications for outer disk ionization.

To address this problem we took a subsample of points throughout the disk and computed a weighted optical depth evenly spaced over 4$\pi$ steradians.  The details of this approach are discussed in Appendix~\ref{app:isrf}.  Unlike the case of stellar FUV ionization, it is not carbon, but rather sulfur that feels the interstellar UV ionization in the outer disk edge.  This difference arises because in the outer disk, CO self-shielding severely limits the thickness of the \ion{C}{2} ionization front.  Sulfur self-shields as well, but at a much less efficient rate \citep{pbc2011b}, and therefore the thickness of the \ion{S}{2} front is set instead by dust attenuation.  Combining the stellar and interstellar UV fields, we compute a simple equilibrium sulfur chemistry (see Appendix~\ref{app:sulf}) to determine the volumetric ionization rate arising from the ISRF.

\begin{figure}
\begin{centering}
\includegraphics[width=2.85in]{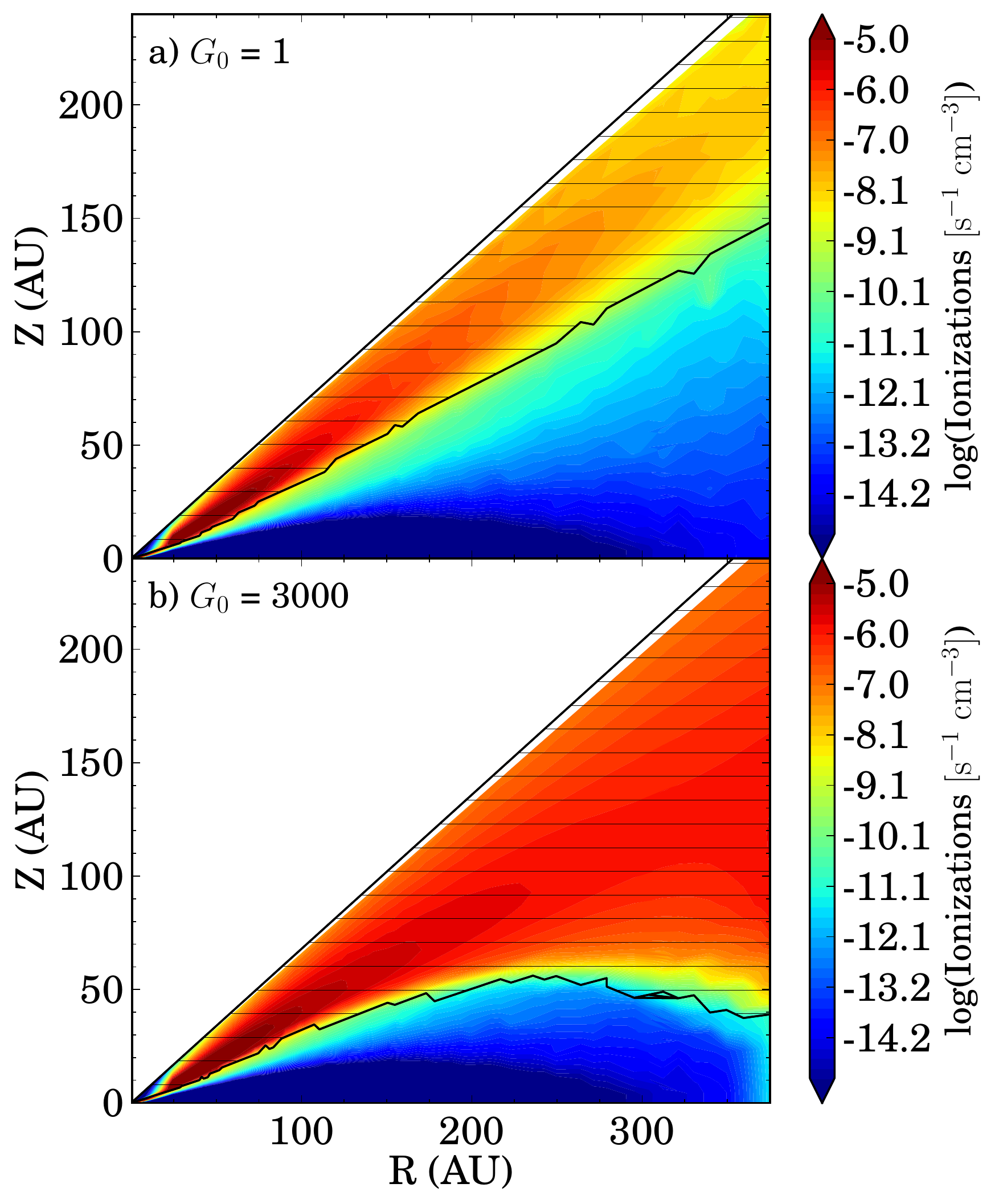}
\caption{Volumetric ionization rate (filled color contours) from stellar FUV and interstellar FUV combined for the cases of: a) a typical external field, $G_0$ = 1, and b) an elevated cluster-like field, $G_0$ = 3000.  The black hatched lines denote the region where the total (stellar plus external) FUV flux provides $>$ 30\% of the ionizing radiation.  The $G_0$ = 1 case is similar to the star-only, while the cluster scenario provides a significant source of ionization both at the surface and outer edge of the disk.}
\label{fig:isrf}
\end{centering}
\end{figure}

We re-plot the combined fractional contribution from stellar FUV and interstellar FUV in Figure~\ref{fig:isrf} (see also Figure~\ref{fig:ion3pan} (a), purple lines).  We consider both an interstellar average case $G_0 = 1$ and an enhanced external field, $G_0 = 3000$. The $G_0 = 1$ case is very similar to the star alone, while the $G_0 = 3000$ case shows that for an elevated interstellar flux the ionization from the FUV can become significant.  We note the exact location of the ISRF boundary depends on one's assumptions, namely the sulfur chemistry, which we have significantly simplified.  Nonetheless, the interstellar radiation field, especially in the cluster environment, can create a thin ionized layer on the disk exterior and enhance the surface ionization (Figure~\ref{fig:isrf}, bottom panel) and can even become the dominant source of ionization in the absence of cosmic rays.

\subsubsection{Cosmic Ray Ionization}
\label{sec:crgen}
Galactic cosmic rays are high energy atomic nuclei, largely protons (87\%) with 12\% alpha particles and the remaining $\sim$ 1\% as heavier atoms.  In our Solar System, galactic CRs are strongly modulated by the solar wind making direct measurements of CR spectrum difficult, especially at low energies \citep[see also][]{nath2012}. The extrasolar ISM CR ionization rate $\zeta_{\rm CR}$ has been studied using various observational and theoretical techniques \citep[e.g., using molecular ion emission such as HCO$^+$ and H$_3^{+}$, see discussion and references in][]{indriolo2012} as well as towards sources spanning vastly different physical environments.  Values as high as $\zeta_{\rm CR} \sim 10^{-15}$ s$^{-1}$ have been measured for diffuse cloud sources \citep{mccall2003,shaw2008} while values as low as $\zeta_{\rm CR} \sim 1 \times 10^{-17}$ s$^{-1}$ have been derived for dense cores \citep[e.g., ][]{caselli1998}.  In Figure~\ref{fig:ion3pan}, we use typical values $\zeta_{\rm CR} = 5 \times 10^{-17}$ s$^{-1}$ $\rm H^{-1}$ and $\Sigma_{\rm CR} =$ 96 g/cm$^{2}$ \citep{umebayashi1981}.  It is important to point out that this attenuation column is significantly higher than both FUV photons (10$^{-3}$ g cm$^{-2}$) and 1 keV X-rays (0.5 g cm$^{-2}$), allowing only cosmic rays to penetrate the densest gas.  Furthermore, the disk is isotropically bombarded by cosmic ray particles, causing a greater volume of gas to be more readily ionized by cosmic rays (see Figure~\ref{fig:ion3pan}). We note that recent measurements and models have revised both the value of the CR ionization rate and functional form of $\zeta_{\rm CR}(N_{\rm H_2})$ (discussed below), but for the purposes of illustration in Figure~\ref{fig:ion3pan} we adopt the standard \citet{umebayashi1981} values. 

One explanation for the range in measured cosmic ray ionization rates is variations in the local supernova frequency and proximity, as well as magnetically controlled propagation within molecular clouds \citep{fatuzzo2006}.
\citet{padovani2009} proposed that the range in measured rates across different types of sources can be reconciled by accounting for the low energy cosmic rays ($E_{\rm CR} < 100$ MeV) and their attenuation with column density.  At high densities, \citet{umebayashi1981} find that cosmic rays attenuate exponentially with a critical mass column of 96 g cm$^{-2}$.  To reproduce observations in both low density (diffuse clouds) and high density regimes (cores), \citet{padovani2009} found a better fit to their numerical calculations by using combined power law and exponential terms with relative contributions depending upon choice of incident spectrum, J$ _{\rm CR,0}(E)$. The full functional form of this expression is given by :
\begin{equation}
\zeta_{\rm CR} = \frac{ \zeta_{{\rm pow,0}} \times \zeta_{{\rm exp,0}}}{ \zeta_{{\rm exp,0}} \left (\frac{N({\rm H_2})}{ 10^{20} {\rm cm^{-2}}} \right )^{\alpha}+\zeta_{{\rm pow,0}} \left [\exp{ \left (\frac{\Sigma}{ \Sigma_0} \right )}-1\right ]},
\label{eq:zeta}
\end{equation} 
which reproduces the power law behavior at low densities and exponential behavior at high densities \citep{padovani2013}. The relation between surface and column density is given by $\Sigma = \mu m_{{\rm p}} N({\rm H_2})$ where $m_{{\rm p}}$ is the proton mass and $\mu = 2.36$.  

\begin{figure}
\begin{centering}
\includegraphics[width=3.1in]{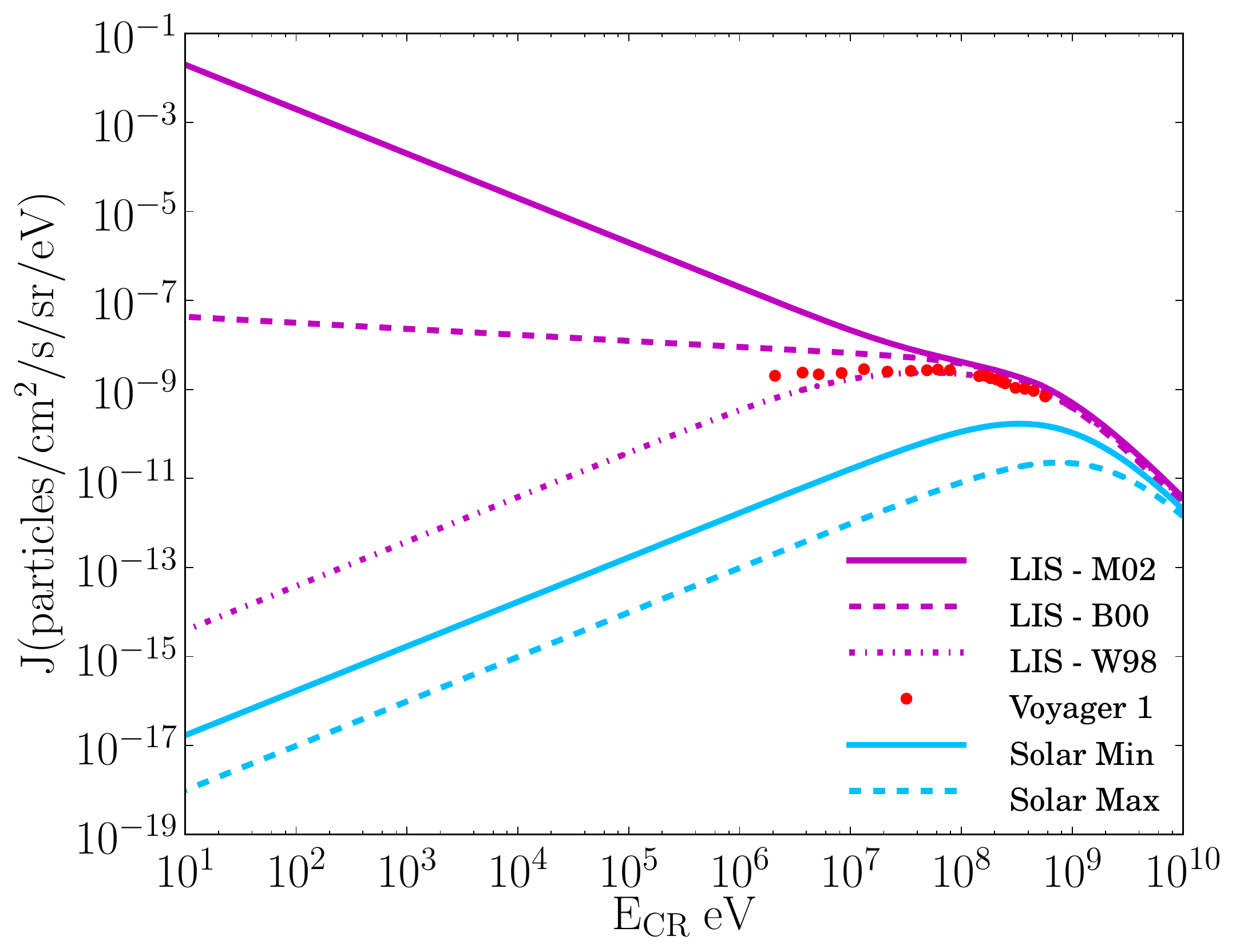}
\caption{Local interstellar (LIS) cosmic ray spectra models (magenta) from top to bottom, solid: \citet{moskalenko2002}, dashed: \citet{burger2000}, dot-dash: \citet{webber1998}.  The red points are the most recent 2013 results from the Voyager 1 spacecraft \citep{webber2013}. Spectra at $R = $ 1 AU (blue) for solar minimum ($\phi = 500$ MeV, solid) and solar maximum ($\phi = 1350$ MeV, dashed). \citet{moskalenko2002} and \citet{webber1998} are the extrapolations of \citet{padovani2009} below $E_{\rm CR} \lesssim 10$ MeV.  \label{fig:solcr}}
\end{centering}
\end{figure}

There are four parameters in this fitting function:  $ \zeta_{\rm pow,0}$, $\zeta_{\rm exp,0}$, $\alpha$ and $ \Sigma_0$.  For a given incident CR spectrum, these parameters describe the shape of the integrated CR ionization rate $\zeta (N_{\rm H_2}) =  \int{4\pi \sigma_{\rm ion} J(E)dE}$.  Because the cosmic ray energy spectrum is only marginally constrained at low energies, \citet{padovani2009} consider two possible extremes for the local interstellar cosmic ray spectra (LIS) shown in Figure~\ref{fig:solcr}.  The first of these is the LIS spectrum determined by \citet{webber1998} [W98] derived from extrapolated {\it Voyager} and {\it Pioneer} measurements up to 60 AU -- generally considered to be the absolute minimum case. The second spectrum is from \citet{moskalenko2002} [M02], which reproduces a large span of supplementary data including the proton, antiproton, and alpha particle spectra as well as the diffuse $\gamma$-ray background.  We note that both spectra have been extrapolated at low energies, $E_{\rm CR} \lesssim 10$ MeV, by \citet{padovani2009} from the original published calculations in the spirit of providing benchmark values for the incident cosmic ray spectra on a molecular cloud.

For each of these LIS spectra (solid and dot-dash magenta lines, Fig.~\ref{fig:solcr}) \citet{padovani2009} then numerically compute and fit \zet\ using the function provided in Equation~(\ref{eq:zeta}).  Table~\ref{tab:zetapar} lists the fitting coefficients for each LIS spectrum (see also Figure~\ref{fig:solcr}). We note that these values have been updated and refit from the original \citet{padovani2009} values \citep[see][]{padovani2013}.

The ISM value of $\zeta_{\rm CR}$ is, however, likely not appropriate for the circumstellar disk environment.  Winds from young stars will be able to shield the disk from cosmic rays at magnitudes that will likely far exceed that of the solar wind due to rapid stellar rotation and strong stellar magnetic fields \citep{svensmark2006,cohen2012}. In the following section we apply the results of \citet{padovani2009} to compute \zet\ incident on the circumstellar disk for various degrees of wind-modulation efficiency, both at solar and more extreme levels.

\section{Exclusion of Cosmic Rays by Stellar Winds}
\label{sec:winds}

\subsection{Modulation By the Solar Wind}
As described in Section~\ref{sec:crgen}, our solar wind drives out low energy ($E \lesssim$ 1 GeV) cosmic rays within a region called the Heliosphere, the very CRs responsible for the bulk ionization of H$_2$ in the ISM.  Furthermore, the degree of modulation by the solar wind varies over the solar magnetic activity cycle by an order of magnitude at energies below $E_{\rm CR} < 100$ MeV (Figure~\ref{fig:solcr}).  Understanding the detailed physics of cosmic ray modulation is still an open question \citep[see review by][]{florinski2011}. To leading order, the dense slow solar wind originates from the hot ($\sim$ 1 -- 2 MK) solar corona and travels at the Sun's escape velocity, carrying with it magnetized plasma frozen-in from the surface of the Sun.  The density of the wind and magnetic field strength decrease with distance from the Sun until the point at which pressure from the ISM overcomes that of the expanding solar wind.  It is here the solar wind is compressed resulting in a magnetic ``pile-up,'' forming a barrier which prevents low energy CRs from freely streaming through the Solar System \citep{weymann1960,burlaga2005,opher2011}.  Young, magnetically active rotating T-Tauri stars are likewise expected to have stellar winds \citep{guenther1997,vidotto2009} in addition to disk winds \citep{hollenbach2000} or X-winds \citep{shu1994}, and therefore it would be unsurprising for T-Tauri stars to similarly drive out low energy cosmic rays within an analogue ``T-Tauriosphere.''  Previous papers have examined reductions in the cosmic ray flux for the early Sun in context of the young, $\lesssim$ 2 Gyr-old Earth \citep{svensmark2006, cohen2012} and find that even at this relatively late stage their models predict substantial reduction in cosmic ray flux.  

It is important to note that even though the mechanism by which the magnetic field is generated on the surface of a T-Tauri star is different from a main sequence dwarf like our Sun, the ability to drive and the properties of a stellar wind simply depends on the presence of a corona, the mass of the star, stellar rotation, and the general magnetic topology on the surface.  Bright X-ray emission from T-Tauri stars is thought to arise from both the stellar corona as well as an accretion shock \citep{kastner2002,brickhouse2010}.  From X-ray measurements, typical temperatures for T-Tauri star's coronas can exceed 10 MK \citep{feigelson1981a,preibisch2005,flaccomio2012} compared to the relatively cooler 1-2 MK solar corona. Such hot coronae are thought to be ``enhanced'' versions of the Sun's corona \citep{feigelson1981a,feigelson1999,favata2003}.  A detailed discussion of the physics behind the link between the solar corona and generation of the solar wind can be found in \citet{gombosi2004}; but in brief, the single most predictive factor of the efficiency of cosmic ray modulation by the solar wind is the magnitude of the magnetic activity at the solar surface.  Between solar minimum and maximum, the cosmic ray flux observed at Earth varies by over an order of magnitude (see Figure~\ref{fig:solcr}).  Additional parameters, like the solar wind speed, the degree of spiral wrapping in the wind, and mass loss rate ($\dot{M}$), are, to leading order, set by the escape velocity of the star, the stellar rotation rate and surface magnetic topology \citep[e.g.,][]{cohen2011}. 

By using our knowledge of how cosmic ray modulation by winds operates in our only measured example, the Solar System, we can begin to learn something about how stellar wind modulation may operate and impact the circumstellar environment in other systems. In the following section we use the results of \citet{usoskin2005} to make simple predictions for scaled-up degrees of cosmic ray modulation.  To determine \zet\ in the extreme case of a T-Tauri star, we compare solar cosmic ray modulation against various cyclical solar parameters relating to the solar magnetic activity.  We then extrapolate these results to obtain a starting point estimate for the degree to wind modulation of cosmic rays operates in the environment of a T-Tauri star.  Such approximations will help illuminate our understanding and interpretation of ionization measurements in disks, which with existing limits on H$_2$D$^+$ towards disks already point to ionization rates lower than ISM \citep{chapillon2011}.

\subsection{The Cosmic Ray Spectrum in a T-Tauriosphere}
\label{sec:incidspec}
While the details of the mechanisms of cosmic ray exclusion are still an active area of research, there is fortunately abundant time-resolved data of the Sun.  Sunspots have been monitored since the time of Galileo, and the Wilcox Solar Observatory has conducted daily observations of the Sun's global magnetic field since 1975\footnote{http://wso.stanford.edu/}.  Space weather is monitored on the timescale of minutes and cosmic ray rates have been monitored hourly (or more frequently) since 1964\footnote{http://cosmicrays.oulu.fi}.  Such a wealth of time sequence data is  useful because properties of the solar wind -- the excluder -- are set largely by the Sun.  The winding of the field is determined by solar rotation; the magnetic field is for the most part frozen in from the Sun's surface.  The magnetic activity cycles on the Sun are imprinted on the solar wind, and in turn directly impact the CR-modulating ability of the wind. 

To empirically relate solar activity and the strength of cosmic ray modulation by stellar winds we use a parametric form of the differential energy spectrum of cosmic rays $J_{\rm CR} (E)$ at 1 AU.  The expression below is commonly known as the ``force-field'' approximation, and it provides a very useful way to describe the observed shape of the cosmic ray spectrum throughout the solar cycle using a single parameter, the modulation potential $\phi$ \citep{usoskin2005} with good astrophysical agreement at heliocentric distances, $D$, near $D$ = 1 AU \citep[e.g.][]{caballerolopez2004,usoskin2005}.  Caveats of this approximation are discussed later in this section.  The modulated cosmic ray proton spectrum $J_{\rm CR} (E)$ in units of cm$^{-2}$ s$^{-1}$ sr$^{-1}$ eV/nucleon$^{-1}$ is:
\begin{equation}
J_{\rm CR} (E,\phi) = J_{{\rm LIS},{\rm CR}} (E+\phi) \frac{E(E+2E_r)}{(E+\phi) (E+\phi+2E_r)},  \label{eq:forcefield}
\end{equation}
where
\begin{equation}
J_{{\rm LIS},{\rm CR}} (E) = \frac{1.9 \times 10^{-9}~P(E)^{-2.78}}{1+0.4866~P(E)^{-2.51}},  \label{eq:forcelis}
\end{equation}
with $P(E) = \sqrt{E(E+2E_r)}$ and the proton rest mass energy $E_r = 0.938$ GeV. In Equations~(\ref{eq:forcefield}) and~(\ref{eq:forcelis}), $E$ is in GeV per nucleon and the modulation potential $\phi$ is in GeV.  We note that values of the modulation potential $\phi$ in the literature and in this work are most frequently given in MeV, but in the commonly used functional form reproduced in Equation~(\ref{eq:forcelis}) energy must be provided in GeV.  The LIS spectrum assumed by \citet{burger2000} [B00], given in Equation~(\ref{eq:forcelis}), is typically used and is shown in Figure~\ref{fig:solcr}. Since this function is used to fit cosmic ray data, we stress that the specific assumption for $J_{\rm LIS}$ does not matter so long as the fit is accurate and it is not changed (it acts as a normalizing factor).  Indeed, according to \citet{usoskin2005} different definitions for $\phi$ have led to confusion in the literature, and in that work the authors attempt to reconcile this confusion and reconstruct a large time baseline of $\phi$ values, looking at the longterm variations in $\phi$ over roughly five solar cycles (see Fig. 7 of that work).  

In addition to variations over the solar cycle, the modulation efficiency of the wind varies with heliocentric distance.  For example, the cosmic ray flux at $E_{\rm CR} =$ 300 MeV varies from $D = 1$ AU to $D = 80 $ AU by a factor of $\sim 6$ \citep{caballerolopez2004}.  We emphasize that the force field approximation is indeed a simple approximation, which tends to {\it over predict} the cosmic ray flux at low energies at large heliocentric distances.  For example, at $D$ = 60 AU, the force field approximation over predicts the 20 MeV CR proton flux by a factor of $\sim 4.2 \times$ as compared to a full numerical model of the one-dimensional cosmic ray transport equation \citep[see Fig. 2 of][]{caballerolopez2004}.  For energies above $E_{\rm CR} \gtrsim 80$ MeV, however, the approximation improves significantly and the predicted differential cosmic ray fluxes are in agreement with the full numerical model to better than 20\% accuracy.  For our simple models we assume a constant modulated spectrum incident on the disk as computed at $D = 1$ AU without radial variation; however, in Section~\ref{sec:totzone}, we consider the effect of a positive cosmic ray flux gradient on disk ionization.

\begin{figure*}
\begin{centering}
\includegraphics[width=6.5in]{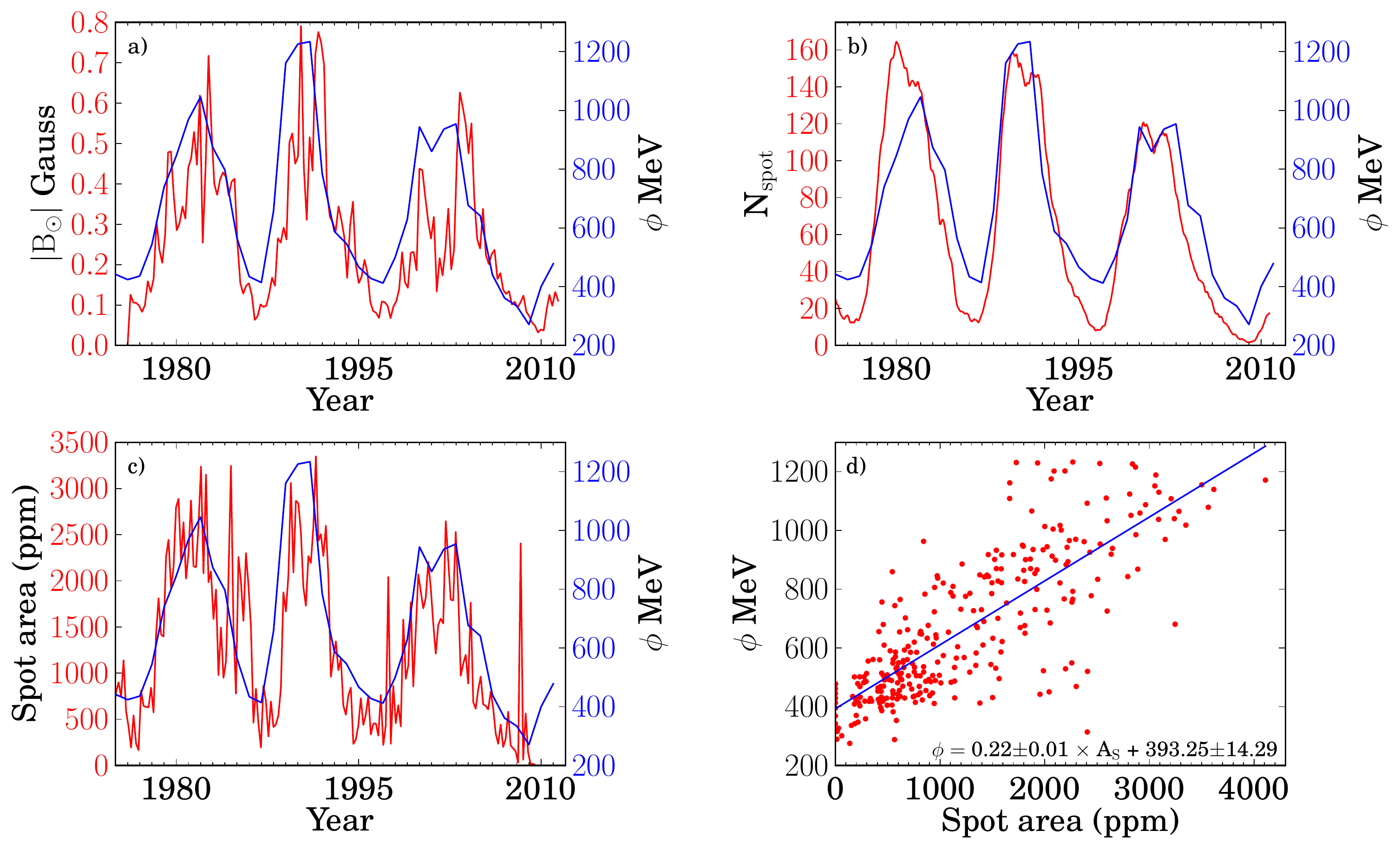}
\caption{Time correlation of the cosmic ray modulation parameter $\phi$ \citep{usoskin2005} with other solar quantities: a) solar mean magnetic field from the Wilcox Solar Observatory, b) number of sunspots \citep[{\it SPIDR};][]{oloughlin2012}, c) fractional area covered by sunspots in millionths of the solar surface area \citep{balmaceda2009}, and d) the correlation between spot area and modulation parameter with a linear fit.  Fit parameters listed at the bottom right.}  \label{fig:solpar}
\end{centering}
\end{figure*}

To attempt to extrapolate the magnitude of modulation from solar values to the case of a more magnetically active T-Tauri star, we have correlated the time aggregated values of the modulation potential $\phi(t)$ \citep{usoskin2005} against other time-resolved measured solar quantities, including mean magnetic field strength on the Sun from the Wilcox Solar Observatory data, number of sunspots from the {\it SPIDR}\footnote{http://spidr.ngdc.noaa.gov/spidr/} database \citep{oloughlin2012}, and fractional area coverage of sunspots \citep{balmaceda2009} shown in Figure~\ref{fig:solpar}.  Because the mass loss rate in the solar wind is related to the coverage of open magnetic field line regions \citep{cohen2011}, tracing the correlation between $\phi$ and the magnitude of the open $\vert B \vert$-field component via solar coronal hole measurements would prove the most useful.  Coronal holes reveal regions where plasma can freely escape along open field lines, in contrast to X-ray bright regions where the hot plasma is trapped.  The time coverage of coronal hole observations, however, cover just over one solar cycle \citep{insley1997}, and therefore the correlation cannot be accurately determined without a longer baseline of data.  In Figure~\ref{fig:solpar} we plot the solar mean magnetic field amplitude, the number of spots and the area of spot coverage alongside $\phi$ (blue curve in panels a, b, and c) as a function of time.  
By linking the state of the solar magnetic activity with the modulation parameter, $\phi$, we can extrapolate $\phi$ to make a simple prediction for the degree of cosmic ray modulation for a more magnetically active young star.  

These quantities are convenient as they can be measured for other stars, specifically magnetic field strength and spot coverage.  Number of spots is less meaningful as T-Tauri stars are suggested to have single spots covering large areas \citep{donati2007,donati2011a,donati2011b,donati2012}. The magnetic field strengths on T-Tauri stars are complex, multicomponent and span a large range in magnitude \citep[see the overview in ][]{johnskrull2007}.  Recent work to map photospheric magnetic topology on a handful of objects using spectropolarimetry \citep[e.g., ][]{donati2003,donati2007} may allow us in the future to link coronal hole coverage to radial field components on T-Tauri stars.  For the time being, we are left with spot coverage area as the proxy for magnetic activity and cosmic ray exclusion.  

\begin{figure}
\begin{centering}
\includegraphics[width=3.15in]{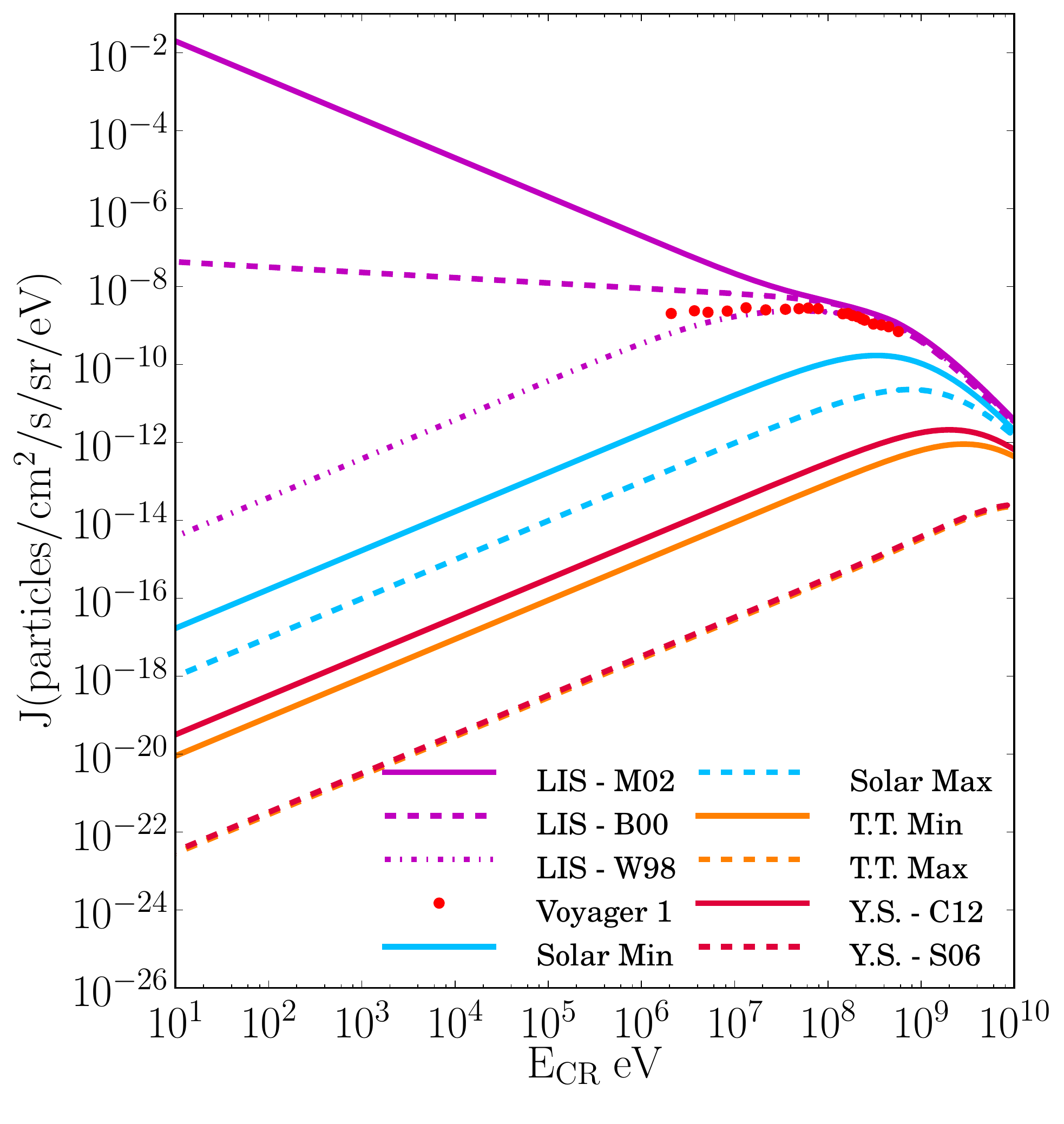}

\caption{Modulated cosmic ray spectra with unmodulated ISM rates shown for comparison (magenta), as well as red points from recent Voyager 1 measurements, same as Figure~\ref{fig:solcr}.  Blue lines -- Solar: minimum $\phi = 500$ MeV and maximum $\phi = 1350$ MeV.  Orange lines -- T-Tauri extrapolation models: 2\% spot coverage ($\phi = 4800$ MeV) orange dashed line and 8\% spot coverage ($\phi = 18$,$000$ MeV).  Solid dark pink line shows a fit by Eq.~(\ref{eq:forcefield}) to the results of \citet{cohen2012} ($\phi = 3500$ MeV) and \citet{svensmark2006} ($\phi =  17$,$500$ MeV) both for the 800 Myr-old Sun.  \label{fig:allcr}}
\end{centering}
\end{figure}

Fractional coverage by spots ranges from 3\% to 17\% and is time variable \citep{bouvier1989}.  Extrapolating the results in Figure~\ref{fig:solpar} (d) yields cosmic ray modulation parameter values of $\phi$ = 4800 MeV, 9200 MeV, and 18,000 MeV for 2\%, 4\% and 8\% spot coverage respectively.  Under the force-field approximation, these modulation parameters, $\phi$, fully describe the shape of the differential cosmic ray energy spectrum, $J_{\rm CR} (E,\phi)$, given in Equation~(\ref{eq:forcefield}).  For spot coverage $ \ge $ 10\% and a stellar X-ray luminosity $L_{\rm XR} \ge 10^{29}$ erg s$^{-1}$, the cosmic ray flux falls below the ionizing X-ray flux from the star and can be neglected throughout the disk (see also Figure~\ref{fig:ion3pan}). Figure~\ref{fig:allcr} shows the incident differential cosmic ray spectra $J_{\rm CR} (E,\phi)$ from Equation~(\ref{eq:forcefield}) with the caveats outlined in Section~\ref{sec:incidspec} for modulation at solar minimum, solar maximum, T-Tauri minimum (2\%), and T-Tauri maximum (8\%).

Using this simple empirical estimate, we find similar magnitudes of CR exclusion as the theoretical models of \citet{cohen2012} [C12] and \citet{svensmark2006} [S06], which predict reduced cosmic ray fluxes at Earth under the conditions present for the young Sun (Y.S.), at age $t = 0.8$ Gyr.  We fit and extrapolate their results using Equation~(\ref{eq:forcefield}) and show these fits in Figure~\ref{fig:allcr} (Y.S., dark pink lines) for comparison. While the specific flux of cosmic rays entering a T-Tauriosphere depends on either simplifying assumptions, i.e., our ``scaled-up'' Heliosphere approach, or the specifics of the detailed models, it is clear that cosmic rays are likely excluded at a substantial degree, at least $\sim 3$ orders of magnitude below solar levels for 1 MeV cosmic rays, and equivalently $\gtrsim 6$ orders of magnitude below the inferred ISM CR flux.

\begin{centering}
\begin{deluxetable}{lcccc}
\tablewidth{0pt}
\tablecolumns{5} 
\tablecaption{$\zeta_{\rm CR}$ fitting parameters for different incident spectra shown in Figure~\ref{fig:allcr}, see Eq.~(\ref{eq:zeta}).   \label{tab:zetapar}}   
\tabletypesize{\footnotesize}
\tablehead{
  \colhead{Model}  & 
  \colhead{$ \zeta_{\rm pow,0}$}   &
  \colhead{$ \alpha$}    &
  \colhead{$\zeta_{\rm exp,0}$} &
  \colhead{$ \Sigma_0$  }      \\            
  \colhead{ }  & 
  \colhead{[s$^{-1}$]}   &
  \colhead{}    &
  \colhead{[s$^{-1}$] } &
  \colhead{[g cm$^{-2}$]}                              
}
\startdata
ISM M02 & 6.8$\times 10^{-16}$ & 0.423 & 3.7 $\times 10^{-18}$ & 210  \\
ISM W98 & 2.0 $\times 10^{-17}$ & 0.021 & 9.4 $\times 10^{-19}$ & 260 \\
Solar Min & 1.3 $\times 10^{-18}$ & 0.00 & 3 $\times 10^{-18}$ & 190 \\
Solar Max & 2 $\times 10^{-19}$ & -0.01 & 8 $\times 10^{-19}$ & 230 \\
T-Tauri Min & 1 $\times 10^{-20}$ & -0.03 & 2 $\times 10^{-19}$ & 270 \\
T-Tauri Max &  3 $\times 10^{-22}$ & -0.03 & 2 $\times 10^{-19}$ & 270 
\enddata
\end{deluxetable}
\end{centering}

\subsection{Extent of the T-Tauriosphere}
How large do we expect the T-Tauriosphere, i.e., the circumstellar region of reduced cosmic ray flux, to be?  The Sun's Heliosphere, for example, extends out to at least $R \sim 120$ AU. The boundary of the Heliosphere, called the Heliopause, is set roughly by the balance of the outward magnetic and ram pressure from the solar wind and the external pressure from the surrounding ISM.  If the T-Tauriosphere only encompasses the inner regions of the disk, the outer disk would be left fully exposed to galactic cosmic rays.  

The external pressure from the ISM in the solar neighborhood is approximately $P_{\rm ISM} \sim B_{\rm ISM}^2/8\pi$, where $B_{\rm ISM} \sim$ 3 -- 10 $\mu$G.  To estimate the outward pressure from a T-Tauri star's stellar wind we must make a few assumptions.  For the internal magnetic and ram pressure $P_{\rm mag}$ and $P_{\rm ram}$ we must assume a wind flow velocity $v$, mass-loss rate $\dot{M}$, and magnetic field dependence $B_{\rm w}$.  The flow velocity is typically of order the star's escape velocity, and the wind's magnetic pressure is typically negligible at all radii compared to the ram pressure (see below). The mass-loss rate is a bit more complicated and depends on whether the wind is a stellar or disk wind.  How a disk wind would interact with the T-Tauriosphere would require a full MHD treatment and is thus beyond the scope of this paper. However, this interaction could be important and could lead to larger regions of CR exclusion.

The outward pressure can be written as $P_{\rm wind} = P_{\rm mag} + P_{\rm ram} = B_{\rm w}(R)^2/8\pi + \dot{M} v /4 \pi R^2 $. The radial dependence of the magnetic field term depends on how tightly wound the wind is; for example, a perfectly radially outward flowing wind would drop as $B \sim R^{-2}$, but more circumferentially wrapped wind would drop as only $R^{-1}$, causing $P_{\rm mag}$ to drop as $R^{-4}$ and $R^{-2}$ respectively.  Moreover, such a highly wrapped wind would exclude galactic cosmic rays even more efficiently, therefore leading to even more severe modulation not considered here.  \citet{sterenborg2011} created a grid of solar wind models to study the physical properties of the stellar wind when the Sun was $\sim$ 1 Gyr-old.  That work found that the stellar wind was typically slower, had higher mass-loss rates and was stronger magnetically than at the present day, with typical values of $v = 266$ km/s, $B = 0.25$ mG, and $\dot{M} = 1.42 \times 10^{-12}$ $M_{\odot}$ yr$^{-1}$ taken at $\sim$ 1 AU respectively.  At 1 AU the contribution from each pressure component is $P_{\rm mag} = 2.3 \times 10^{-9}$ dyne cm$^{-2}$ and $P_{\rm ram} = 8.4 \times 10^{-7}$ dyne cm$^{-2}$; as a result, for either radial or tightly would wind magnetic fields, $P_{\rm ram}$ always exceeds $P_{\rm mag}$ outside of $\gtrsim$ 0.1 AU.

Therefore, using the simple equality $P_{\rm ISM} = P_{\rm wind} \rightarrow B_{\rm ISM}^2/8\pi = \dot{M} v /4 \pi R^2$, we can solve for the radius at which the external ISM pressure and outward wind pressure balance.  Given the wind values above, the boundary occurs at 1540 AU and 460 AU for $B_{\rm ISM} =$ 3 $\mu$G and 10 $\mu$G respectively.  The values of the parameters discussed above are applicable for a very young Sun, but are old compared to the age of a T-Tauri star, 1-10 Myr.  Youthful T-Tauri stars are even more magnetically active and likely have elevated coronal activity, resulting in much higher stellar-wind mass loss. Furthermore, rapid rotation characteristic of the first $\sim$ 30 - 50 Myr will strongly enhance the stellar wind's ability to modulate the cosmic ray flux \citep[see Fig. 2 of ][]{cohen2012}.  This can operate in tandem with mass loss from a disk wind, creating substantially higher mass loss rates $\dot{M}$.  Based upon observed T-Tauri X-ray fluxes, \citet{decampli1981} predict that a T-Tauri star's hot coronal gas has sufficient pressure to power mass loss rates via a stellar wind of up to $\dot{M} \sim 10^{-9} M_{\odot}$ yr$^{-1}$.  With all other parameters held constant, this would correspond to an $R \sim $ 12,000 AU-sized T-Tauriosphere for a $B_{\rm ISM} =$ 10 $\mu$G background magnetic field, much larger than typical disk sizes of a few hundred AU.  At this point the outer boundary could instead be set by intervening remnants of the parent molecular cloud.

How the disk and environment would interact with the solar wind ``fluid'' however is beyond the scope of this paper, e.g., is the T-Tauriosphere doubly-lobed or does it flow over the disk producing wind eddies and vortices?  How do the magnetic fields contained in the stellar winds and disk winds connect with external magnetic fields at an analog ``T-Tauriopause?''  Nonetheless, as we have demonstrated, the region of cosmic ray exclusion likely extends over a large region and it would not be unreasonable for it to fully enclose the disk, even for massive disks hundreds of astronomical units in radius, and therefore for the remainder of this paper we consider the incident cosmic ray spectra to be uniformly modulated over the entire disk.

\subsection{Cosmic Ray Attenuation: $\zeta_{\rm CR} (N_{\rm H_2})$}\label{sec:zetanh}
From these cosmic ray spectra $J_{\rm CR} (E)$ described in Section~\ref{sec:incidspec} we can now determine the integrated ionization rate as a function of H$_2$ column density. The study of cosmic ray penetration clouds and circumstellar disks has an extensive history \citep[e.g.][]{hayakawa1961,cesarsky1978,umebayashi1981,padovani2009} The interaction and attenuation of a cosmic ray in molecular matter critically depends on its initial energy. As a result the shape and energy range of the incident CR intensity spectrum directly determines the integrated ionization rate $\zeta =  \int{4\pi \sigma_{ion}(E) J_0(E)dE}$, where $J_0(E)$ is the differential cosmic ray spectrum (from now on simply spectrum) at the disk surface (see Fig.~\ref{fig:allcr}) and $\sigma_{ion}(E)$ is the energy dependent ionization cross section of H$_2$. 

\begin{figure}
\begin{centering}
\includegraphics[width=3.20in]{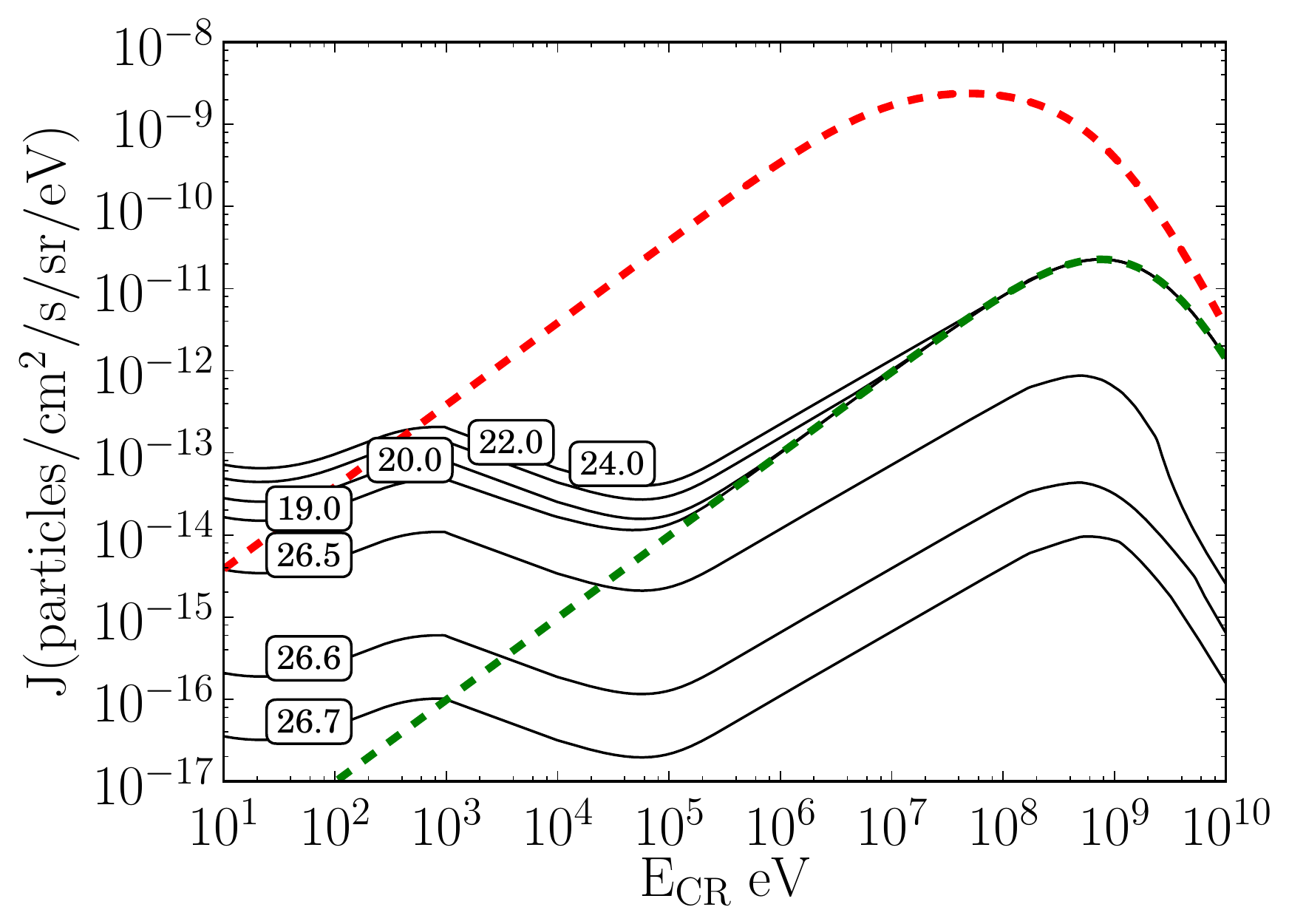}
\caption{Cosmic ray spectra as a function of column density $N_{\rm H_2}$ cm$^{-2}$ for the case of a solar maximum incident spectrum $\phi = $ 1350 MeV. Note the initial rise at low energies due to the reprocessing of high energy cosmic rays.  Box-labels denote log(${N_{\rm H_2}}/{\rm cm^{-2}}$), the dashed green line is the incident spectrum, and the W98 spectrum is shown in red for comparison. \label{fig:diffspec_nh}}
\end{centering}
\end{figure}

Here we compute the ionization rate $\zeta_{\rm CR}$ as a function of depth, given a ``heliospheric'' incident CR proton spectrum on a disk.  In general, the decay of cosmic ray flux with column density can be thought of as an energy reprocessing of CR particles with column density.  To compute the density evolution of the cosmic ray flux, we follow the ``continuous slowing down approximation'' method of \citet{padovani2009} (hereafter P09). Ionization processes included are ionizations of H$_2$ and He by protons and electrons, electron capture, dissociative ionization, double-ionization, and a correction for ionization by secondary particles.
We consistently solve for the particle travel range, Equation (21) of that work, the energy loss incurred, and subsequent particle energy reprocessing, see Equation (25) of P09.  Furthermore, we include a correction for secondary ionizations following \citet{glassgold1973} with a logarithmic extrapolation at high energies as was done in P09.  In Figure~\ref{fig:diffspec_nh} we show the resulting spectra for the case of solar maximum wind modulation ($\phi = 1350$ MeV) at the indicated log column densities.  This can be directly compared with the column density evolution presented in Figures 9 and 10 of P09 for the M02 and W98 incident spectra.  We note that from this point, we adopt the M02 spectra as our ``true'' interstellar spectrum and compare our modulated results to this.

\begin{figure}
\begin{centering}
\includegraphics[width=3.2in]{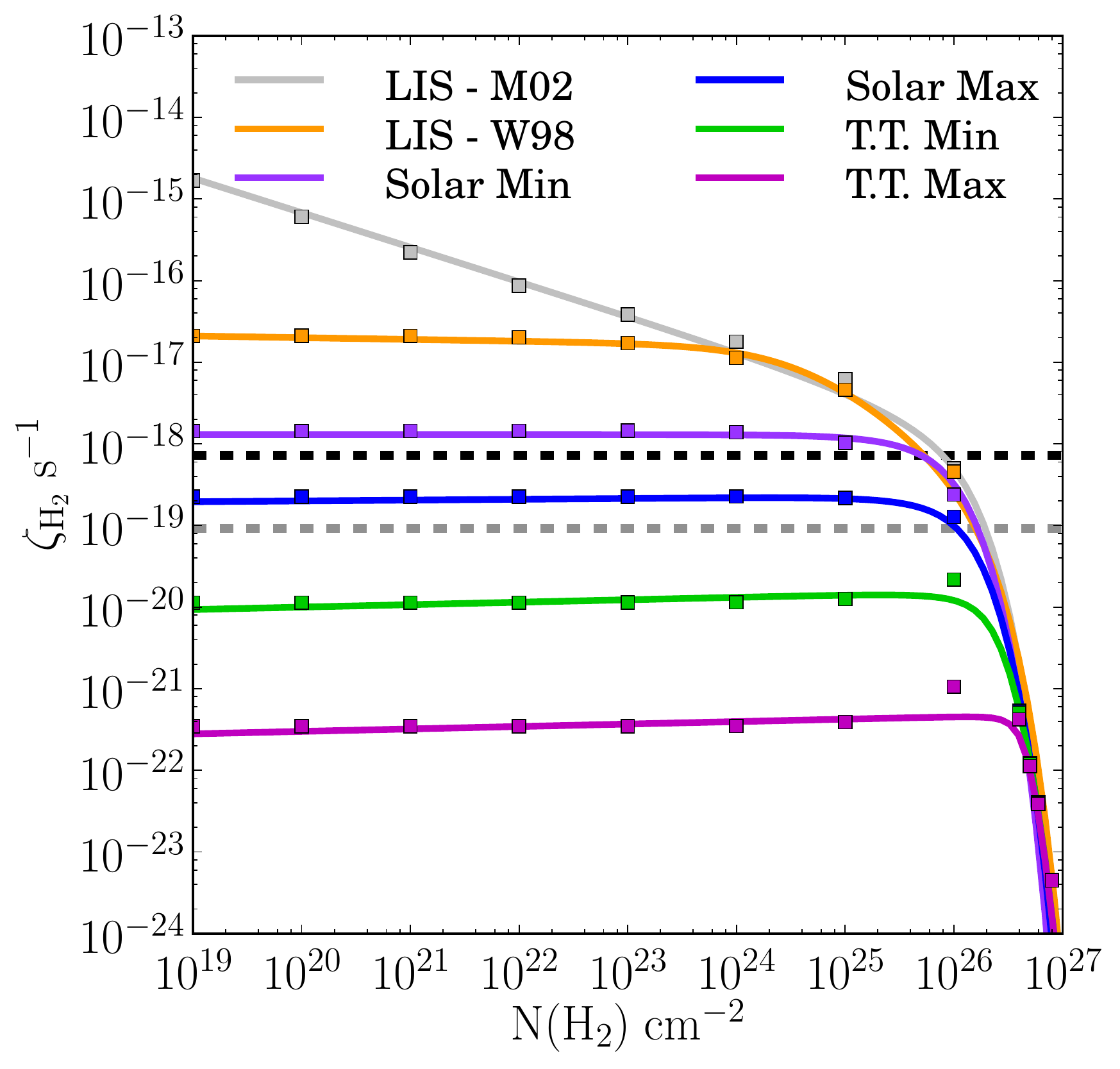}
\caption{Integrated cosmic ray ionization rates as a function of column density for different incident CR spectra.  Note: the solar values and extrapolated T-Tauri models are formally taken at $D = 1$ AU.  Squares are the computed values while solid lines show the best fit to Eq.~(\ref{eq:zeta}).  Fitting parameters are given in Table~\ref{tab:zetapar}, and labels are as described in Figures~\ref{fig:solcr} and \ref{fig:allcr}.  For comparison, the black dashed line represents the inferred ionization rate due to radionuclide decay in the Solar Nebula while the gray dashed line is derived from the mean interstellar abundance of $^{26}$Al \citep{umebayashi2009}. \label{fig:zetanh}}
\end{centering}
\end{figure}

Finally, each differential energy curve $J(E_{\rm CR})$ is then integrated between $E_{\rm CR}$ = I(H$_2$) and 100 GeV to produce a total ionization rate per H$_2$: $\zeta(N_{\rm H_2}) =  \int{4\pi \sigma_{ion}(E) J(E)dE}$ s$^{-1}$, where $\sigma_{ion}$ is the CR ionization cross section provided by P09.  These results are shown in Figure~\ref{fig:zetanh} (squares).  These points are then fit (solid lines) using the function provided in Equation~(\ref{eq:zeta}) and the corresponding coefficients are listed in Table~\ref{tab:zetapar}.  For the most severely modulated T-Tauri models there is a small rise in the ionization rate at high column density that results from particle conservation of the reprocessed high energy particles.  This effect is otherwise hidden in the models that have more initial cosmic rays energies below $E_{\rm CR} < 100$ MeV.  We do not fit the bump using the simple parameterization as it only deviates from the otherwise good fit by less than a factor of $\sim$ 2.

As can be seen, modulation of cosmic rays even by a solar-like wind has a significant effect on the integrated cosmic ray ionization rate, $\zeta_{\rm CR}$. While the LIS ionization rates vary between $\zeta_{\rm CR} \sim 10^{-15} - 10^{-17}$ s$^{-1}$, the effect of solar minimum modulation on $\zeta_{\rm CR}$ is more than an order of magnitude below the LIS values, and the {\it unattenuated} incident CR rates from our simple extrapolation model for the 2\% and 8\% spot-covered T-Tauri stars are just $\zeta_{\rm CR} \sim 10^{-20}$ s$^{-1}$ and $\sim 3 \times 10^{-22}$ s$^{-1}$, respectively.  Below these values, scattered stellar X-ray ionization of H$_2$ begins to dominate the weak cosmic ray field.  Indeed, the cosmic ray flux in these cases is so low that, if present, ionization by decay products of short-lived radionuclides ($\zeta_{\rm RN} = 7.3 \times 10^{-19}$ s$^{-1}$) as inferred from the early Solar Nebula \citep{umebayashi2009} can readily dominate the cosmic ray and X-ray ionizing flux in the disk midplane (dashed black line, Figure~\ref{fig:zetanh}); see also Section~\ref{sec:radnuc} for further discussion.

\section{Exclusion of Cosmic Rays by Large Scale Magnetic Fields}
\label{sec:mag}

Cosmic rays are intrinsically high velocity ions and as such their trajectory will be shaped by the presence of a magnetic field. In the protoplanetary disk environment magnetic structure may originate from the central star and/or remain from earlier stages in protostellar development, i.e., from the collapse of the parent cloud.  The CR's motion will be directed by the field lines so long as the gyroradius, $r_g$, of the ion is less than the relevant scales considered, where the smallest scale of interest is $\sim R_{{\rm disk}}$.  For a cosmic ray proton of energy $E_p = 1 $ TeV and a magnetic field strength $B = 10 \mu$G, the gyroradius $r_g = \gamma m v_{\perp} c /(ZeB) \sim 0.02 $ AU, much smaller than $R_{\rm disk} \sim $ 100 -- 1000 AU.  Given that the field strength is expected to be generally higher than this value and the energy of cosmic rays lower, we are safely within the bounds of this criterion for all particles considered.

In the following sections we introduce the magnetic topologies considered, how mirroring and funneling modulate the propagation of galactic cosmic rays, and how this modulation impacts the disk ionization state. 

\subsection{Environmental Magnetic Fields}

Dust polarimetric observations of young protostars in some instances exhibit large scale magnetic field structure in what has been described to resemble the shape of an hourglass \citep[e.g.,][]{girart2006}, thought to arise from the gravitational collapse of a magnetized molecular cloud \citep[see review by][]{crutcher2012}.  The large scale magnetic structure surrounding a T-Tauri star is difficult to constrain, however. T-Tauri stars are known to be strongly magnetized \citep[e.g., ][]{basri1992,johnskrull1999,yang2005}; how and if the stellar fields couple to a large scale background is unknown.  However, by necessity the disk must be magnetized for MRI to initiate, and while deep within the disk the magnetic fields may be randomized by turbulent motions, at large scales the imprint of a protostellar field may still exist.  

For the case of an hourglass-like background magnetic fields, we have solved the semi-analytical magnetized singular isothermal toroid models of \citet{li1996} for a range of mass to flux ratios (i.e., degrees of pinching at the waist of the hourglass), $\lambda  = M/\Phi$. These types of models are representative of the magnetic field of a protostar, and thus provide a first approximation for the ``fossil'' background remnant field near a very young T-Tauri star.  We treat the field as temporally static, with the waist of the magnetic field tied to the disk and at large stellar distances tied to the natal cloud.   

Each model is formally characterized by a parameter $H_0$ \citep[see][]{li1996}, which in practice sets the enhanced magnetic field density in the midplane relative to the cloud. 
In Table~\ref{tab:mass2flux} we provide $H_0$, mass to flux ratios, and the vertical magnetic field strength $B_z$ in the disk midplane at $R = $ 100 AU for each field model.  We note that these mass to flux ratios are related to masses of natal core material over the magnetic flux contained in the core, and not disk masses.  

\begin{deluxetable}{lrrr}
\tablecolumns{4} 
\tablewidth{0pt}
\tablecaption{Hourglass Magnetic Field Parameters.  \label{tab:mass2flux}}   
\centering
\tabletypesize{\footnotesize}
\tablehead{
  \colhead{$H_0$}  & 
  \colhead{$ \lambda $}   &
  \colhead{$ B_{\rm z,100AU}$ [G]} 
  }
\startdata
 2.50e-05 & 852.52 & 7.76e-06 \\ 
1.25e-04 & 387.29 & 1.63e-05 \\ 
 2.50e-04 & 270.26 & 2.42e-05 \\ 
 1.25e-03 & 121.21  & 5.25e-05 \\ 
 6.25e-02 &  13.58 & 3.86e-04 \\ 
 5.00e-01 &   2.66 & 1.43e-03 
\enddata
\tablecomments{Col.~(1): Dimensionless parameter representing the degree of pinching at the waist of a magnetic flux tube.  Col.~(2): Tabulated mass to flux ratio contained in a flux tube $\Phi$ where $\lambda = M(\Phi)/\Phi$.  See \citet{li1996} for further details regarding these parameters.  Col.~(4):  Vertical magnetic field strength at $R = 100$ AU in the midplane.}
\end{deluxetable}

\subsection{Stellar Magnetic Fields}
Zeeman broadening observations of young T-Tauri stars have revealed strong magnetic fields at the stellar surface with magnitudes of order 100 G - 1.6 kG \citep[e.g.,][]{johnskrull2007}.  The magnetic topology as determined by sensitive spectro-polarimetric measurements is a complex superposition of octupolar, dipolar and split-monopolar magnetic fields \citep{donati2011a,donati2011b,donati2012}.  Higher order field components drop off rapidly at large distances from the star; for example, the dipolar field drops off as $B_{\rm dip} \propto r^{-3}$, and therefore its influence in determining the fate of CRs would only matter very close to the star. Some fraction of the field lines are opened up by the stellar wind to form a split-monopolar configuration, forming an approximately radial field component and hence dropping less steeply as $B_{\rm mono} \propto r^{-2}$.  As a result, the stellar field component that matters most in magnetically directing cosmic ray motion is that contained within the split-monopole component.  

For the stellar field we set the total magnetic field strength at the surface of the star to be $B_{\rm surf} = 3 $ kG and assign a fraction $\gamma$ to be in the radial component.  We then vary the strength of the split-monopolar field as a fraction of the total field. 
In Equation~(\ref{eq:fraccr}) we can simply replace $B_{\rm disk}$ with $B(R) = \gamma B_{\rm tot} (R/R_{\rm star})^{-2}$ where $\gamma$ is the fraction of the stellar magnetic field contained within the split-monopole.  

\subsection{Cosmic Ray Exclusion by Magnetic Mirroring} 
\label{sec:mirror}
\begin{figure}
\begin{centering}
\includegraphics[width=2.1in]{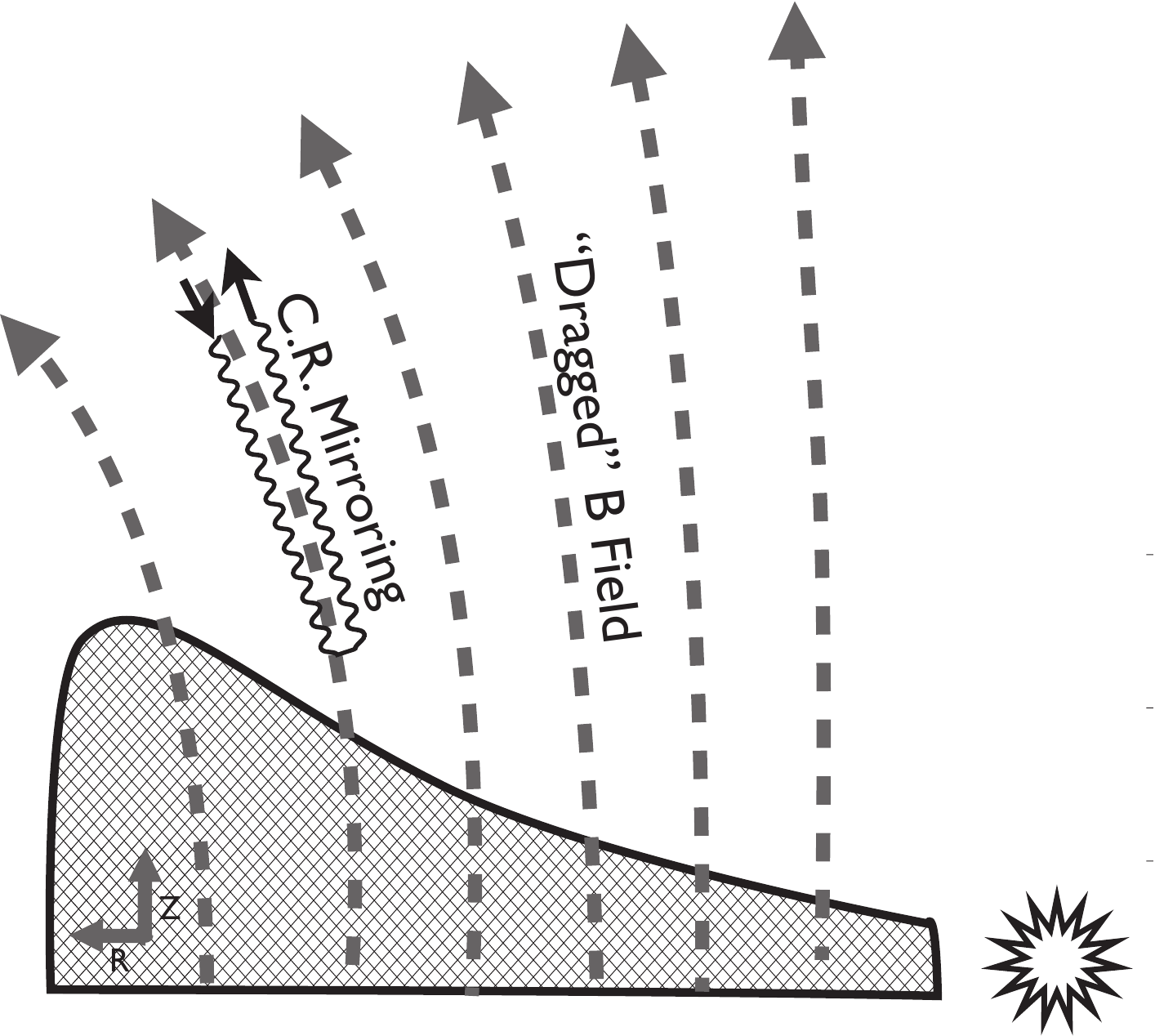}
\caption{Illustration of cosmic ray exclusion by magnetic mirroring.  Dashed lines denote background hourglass-shaped magnetic field and the hatched region indicates the disk, viewed edge on.  Cosmic ray (zig-zag arrow) enters along the magnetic field line but is repelled before reaching the surface of the disk. \label{fig:mirrdia}}
\end{centering}
\end{figure}
The shape of the environmental magnetic field can modulate the propagation of cosmic rays through a process known as magnetic mirroring. The basic principle of mirroring is described as follows \citep[see also ][]{desch2004}.  Charged particles gyrating about magnetic field lines will conserve their total kinetic energy $\propto v_{\parallel}^2 + v_{\perp}^2$ and magnetic moment $\propto v_{\perp}^2/B$.  As the particle enters an area of high magnetic field density it must increase its perpendicular velocity.  As $v_{\perp}$ increases, $v_{\parallel}$ must decrease to keep the total kinetic energy constant.  If the field is pinched to sufficiently high magnetic field strengths, the particle's parallel velocity can be halted ($v_{\parallel} = 0$) and reversed, thus driving the cosmic ray in the opposite direction along the field line.  The cosmic ray thus reflects off of the region of large magnetic field strength (hence the term mirroring).

If $\alpha_{\rm ISM}$ is the initial pitch angle between a CR's velocity vector and the magnetic field, cosmic rays with small $\alpha_{\rm ISM}$ will tend not to mirror (whereas cosmic rays with initial pitch angle $\alpha_{ISM}$ = 90 degrees would
gyrate around the field, although that case represents a set of measure zero). The pitch angle $\alpha$ of a cosmic ray starting with $B_{\rm ISM}$ and $\alpha_{\rm ISM}$ at any given point along the field line is given by
\begin{equation}
\frac{\sin^2{\alpha}}{\sin^2{\alpha_{\rm ISM}}} =\frac{B}{B_{\rm ISM}} = \chi. \label{eq:alpha}
\end{equation}
For a disk threaded with a field of magnitude $B_{\rm disk}$, there is a critical initial pitch angle such that cosmic rays with $\alpha > \alpha_{\rm crit}$ will be repelled before reaching the disk surface; in other words, particles attain $\alpha = 90^{\circ}$ at $B = B_{\rm disk}$ and reverse course.  For particles arriving on one side of the disk, this condition corresponds to a fractional reduction of cosmic rays by mirroring $f_{\rm mirror} = \sqrt{1-[B_{\rm ism}/B_{\rm disk}]^2}$.  

However, simultaneously the open magnetic field lines tend to draw in a larger number of cosmic rays via funneling, enhancing the cosmic ray flux proportionally to the increase in field line density, or $f_{\rm funnel} = B_{\rm ISM}/B_{\rm disk}$.  In general mirroring dominates over funneling \citep{desch2004,padovani2011}, but the effects are of similar magnitude.  The combined fractional removal of cosmic rays as given by \citet{desch2004} is:
\begin{equation}
f_{\rm CR, net} = \left[ \chi - \left(\chi^2-\chi \right)^{1/2} \right], \label{eq:fraccr}
\end{equation}
where $\chi$ is given by Equation~(\ref{eq:alpha}).  In the limit that the magnetic field in \Bdisk\ $\gg$ \Bism, then the net fraction of cosmic rays removed is $f_{\rm CR, net}$ = 0.5.  

For completeness we note that mirroring is only important when the
change in pitch angle due to magnetic field variations is greater than
that due to scattering. The T Tauri systems of interest here are
expected to have large gradients in magnetic field strength, so that
mirroring can be important.  Nonetheless, scattering effects should be
considered in future work.

\begin{figure}
\begin{centering}
\includegraphics[width=2.75in]{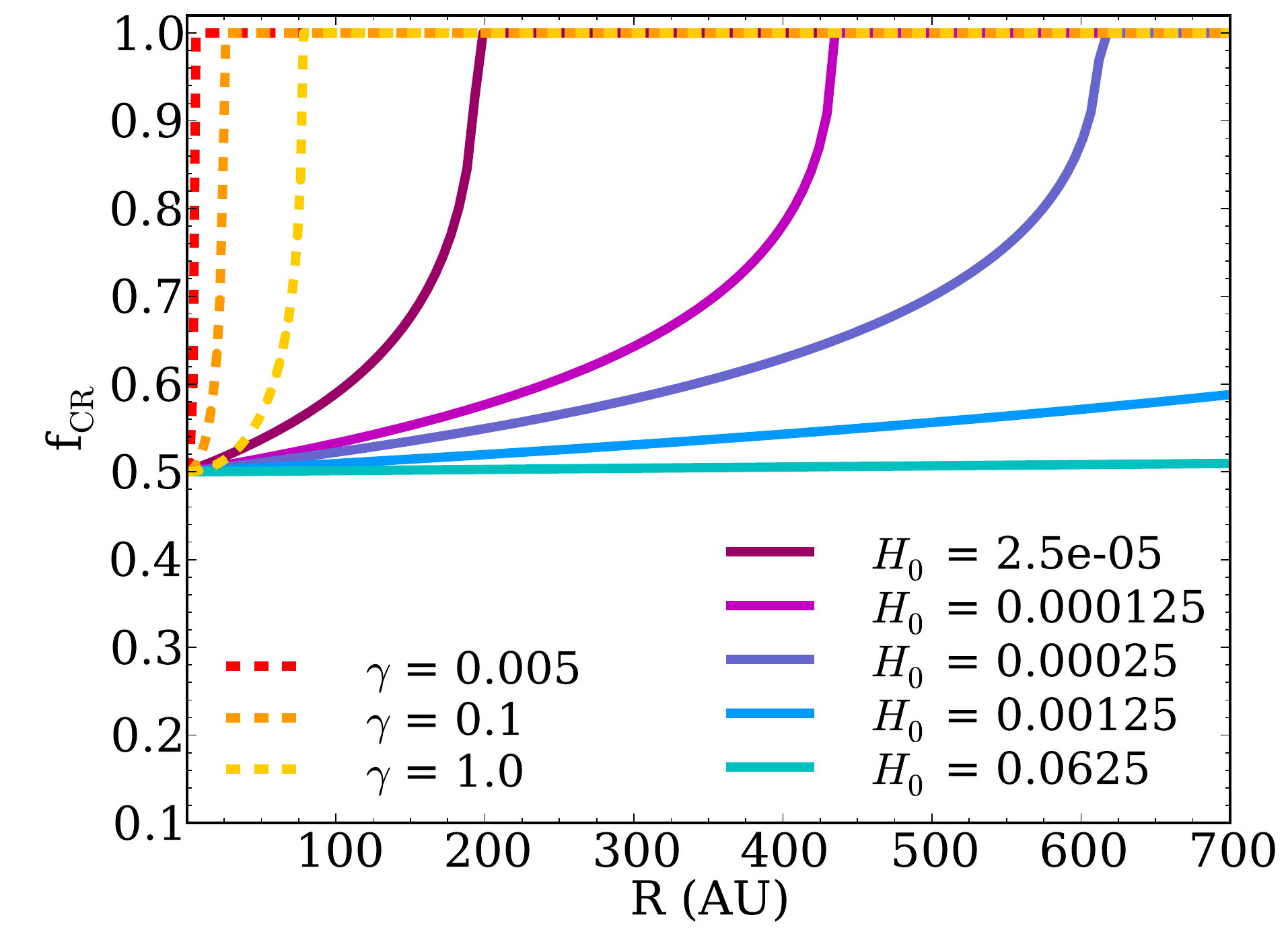}
\caption{Fraction of cosmic rays removed versus disk radius for both the stellar split-monopole models (left, dashed lines) and hourglass models (right, solid lines). \label{fig:mirror}}
\end{centering}
\end{figure}

In Figure~\ref{fig:mirror}, the results for the environmental hourglass model are shown in solid lines and the results for the stellar split-monopolar field are shown in dashed lines. Stellar magnetic effects are only able to modulate cosmic rays relatively near ($R < 100$ AU) the star.  Examples of observed split-monopolar field strengths are 4--5 kG \citep[GQ Lup;][]{donati2012}, 170 G \citep[V4046 Sgr;][]{donati2011b}, and thus we expect stellar magnetic  fields to modulate cosmic ray propagation within 96 AU and 20 AU respectively.  Hourglass magnetic fields can, however, reduce incident cosmic ray fluxes on the scale of hundreds of AU.  For fields that are only moderately pinched at $H_0 \ge$ 0.0625 the entire cosmic ray rate would be reduced by a factor of two.  Such a field configuration corresponds to $B_z = 0.9$ mG at 100 AU in the midplane (see Table~\ref{tab:mass2flux}).

While the magnitude of magnetic modulation is far smaller than the orders of magnitude achieved by a stellar wind, the effect of mirroring more importantly is that this fraction will {\it radially} vary from $f_{\rm CR, net}$ = 0.5 in the inner disk to 1 in the outer disk.  As described in Section~\ref{sec:incidspec} the cosmic ray fluxes will formally experience an energy dependent radial gradient in wind modulation efficiency that can vary by a factor of a few between distances of $D$ = 1 AU to 80 AU \citep{caballerolopez2004,langner2005,manuel2011,manuel2011b}, see also Section~\ref{sec:totzone}. Therefore looking for radial gradients in the disks of T-Tauri stars to learn about extrasolar Heliospheres may be confused observationally with magnetic effects from the star or environment, especially if they are of the same magnitude (i.e., a factor of two). 

\section{Cosmic Ray Exclusion and Dead Zones}
\label{sec:deadz}
The magnitude of cosmic ray exclusion by stellar winds and magnetic fields presented here will have significant implications for the ionization state in the disk.  This in turn will have important implications for the size of the region in disks that is dead to MRI as well as chemical implications to be discussed in Paper II.  In this section we use criteria from simulations of MRI in the literature to compute the size of the active and turbulence-dead regions of the disk and explore how they depend on the cosmic ray ionization rate.

\subsection{Ionization: Active Disk Criteria}
\label{sec:active}
From these ionization rates, $\zeta$, we can determine the electron abundance, which is used to identify the turbulence dead versus active regions. To determine the electron abundance $\chi_e$ we use the steady state expression
\begin{equation}
\chi_e = \sqrt{\frac{\zeta_{{\rm H_2}}}{n \alpha}},  \label{eq:xe}
\end{equation} 
where $\alpha = 2 \times 10^{-6}~T^{-1/2}$ cm$^3$ s$^{-1}$ is the rate coefficient for recombination with molecular ions \citep{glassgold1997}.  This is of course a simple estimate for the ionization state in the gas and does not, for example, include the possibility of charged dust grains.  A more detailed estimate of the ionization fraction including chemistry and charge exchange will be addressed in Paper II.
For the disk to be MRI active, the ions must first be well-coupled to the neutral gas and must have a sufficiently high magnetic Reynolds number, $Re$, given by
\begin{equation}
Re \equiv \frac{c_sh}{D} \approx 1 \left ( \frac{\chi_e}{10^{-13}} \right ) \left ( \frac{T}{{\rm 100 K}} \right )^{1/2} \left ( \frac{a}{{\rm AU}} \right )^{3/2} \label{eq:Re}
\end{equation} where $c_s$ is the sound speed, $h$ is the disk scale height, and $D$ is the magnetic diffusivity \citep{pbc2011a}. 
Recent models by \citet{flock2012} indicate that values of $Re \sim 3300-5000$ are required for sustained turbulence, with a critical value $Re_{\rm crit} \sim 3000$.  We adopt this critical value for the minimum $Re > Re_{\rm crit}$ required for the disk to be MRI active.  

The second criterion for the disk to be unstable to MRI is that there must be frequent ion--neutral collisions for the ions to transfer turbulent motions to the largely neutral disk.  This condition is determined by the ion-neutral collision rate normalized to the orbital frequency,
\begin{eqnarray} 
Am & \equiv & \frac{\chi_i n_{{\rm H_2}} \beta_{in}}{\Omega} \nonumber \\
& \approx &1 \left ( \frac{\chi_i}{10^{-8}} \right ) \left ( \frac{n_{{\rm H_2}}}{{\rm 10^{10} cm^{-3}}} \right ) \left ( \frac{a}{{\rm AU}} \right )^{3/2}, \label{eq:Am}
\end{eqnarray}  \citep{pbc2011a}. 
Simulations by \citet{hawley1998} find that $Am \sim 10^2$ is required for sufficient coupling. However, even in the case of our most strongly ionized models we reach values of at most $\lesssim 3$.  Similarly, \citet{pbc2011a} found even in their most MRI favorable model they could only attain $Am \lesssim 10$ (see \S 4 of that work).  In the weakly ionized limit, \citet{baistone2011} show that a disk can become MRI unstable at {\it any} value of $Am$, as long as the disk is weakly magnetic.  The maximum magnetic field strength such that $Am$ is not a limiting factor is dependent on the ratio of the gas pressure to magnetic pressure, $\beta$.  To be in this regime, $\beta$ must be larger than $\beta_{\rm min}$ (see Eq. (26) of \citet{baistone2011}).  For our disk model this condition is equivalent to $\rm \vert B_{max}(50 AU) \vert \sim$ 0.3 mG -- 5 mG for our most weakly and most strongly ionized models, respectively.   

The magnetic field strength in disks is observationally unknown, however.  Zeeman measurements of molecular clouds give us a ``starting value'' with line of sight estimates of 0.1-1 mG \citep{crutcher2010,falgarone2008}. The fields in disks should thus be at least this strong and likely stronger, amplified by the collapse of the cloud during formation of a protostar.  Therefore, the $Am$ criterion may still be important due to the inferred relatively high magnetic field strengths.  In this work we take $Am > 0.1$ as the critical value for sufficient ion--neutral collisional frequency \citep{baistone2011}.  For completeness, however, in the dead zone plots presented, we indicate both the region that satisfies simultaneously $Am$ and $Re$ (white cross hatched $=$ active) as well as the region that satisfies $Re$ only (outside of orange contour = active); see, for example, Figure~\ref{fig:winddead}.

\subsection{T-Tauriospheric Dead Zones}
\label{sec:totzone}
\begin{figure*}
\begin{centering}
\includegraphics[width=6.1in]{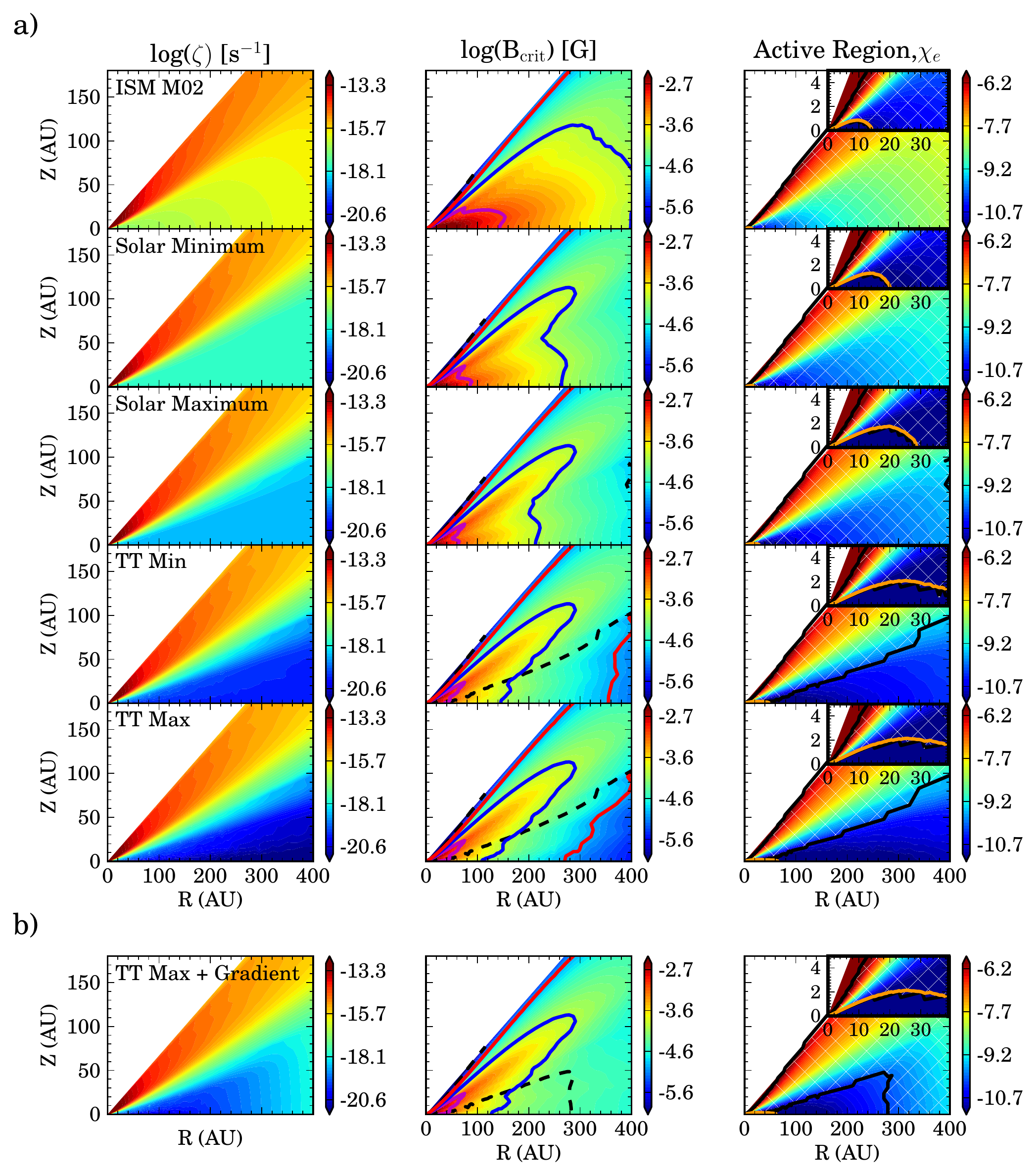} 
\caption{H$_2$ ionization rate $\zeta_{\rm tot} = \zeta_{\rm CR}+\zeta_{\rm XR}$ s$^{-1}$ and corresponding MRI-active regions. a) Uniformly wind-modulated cosmic ray ionization models.  Left column: total H$_2$ ionization rate from stellar X-rays and cosmic rays for $L_{\rm XR} = 10^{-29.5}$ erg s$^{-1}$ and cosmic ray rate as labeled (see Section~\ref{sec:incidspec}). Middle column: critical magnetic field to determine significance of $Am$ criterion (Section~\ref{sec:active}).  Magenta, blue and red lines denote 1 mG, 100 $\mu$G, and 10 $\mu$G respectively.  Right column: MRI-active regions for the various cosmic ray rates.  Regions of the disk that satisfy both the $Re$ and $Am$ criteria are indicated by the white crosshatched region.  The minimum region of the disk that is inactive to MRI, i.e., satisfies $Re$ but not $Am$, is outlined in orange. b) Quantities same as outlined above but now including a 2\%/AU positive radial gradient in the total cosmic ray flux.  \label{fig:winddead}} 
\end{centering}
\end{figure*}

As can be seen in Figures~\ref{fig:allcr} and~\ref{fig:zetanh} the presence of a T-Tauriosphere plays an important role in determining the ionization rate from cosmic rays.  This ionization rate will in turn impact the steady state electron abundances (see Equation~(\ref{eq:xe})) and thus the region of the disk that is dead to MRI turbulence.  In Figure~\ref{fig:winddead} we plot the results for our standard model, varying the cosmic ray flux for each spectrum considered.  In these models we assume the cosmic rays come from both sides of the disk and therefore reduce the cosmic ray contribution per side by half. We integrate the column of material from the surface downward and bottom upward and sum the contributions from both sides of the disk.

The contour plot in the left column shows the total H$_2$ ionization rate from cosmic rays and X-rays combined.  The second column shows contour plots of the critical magnetic field, which sets the relative importance of the $Am$ parameter in determining the disk's dead region.  The third column shows the net result, with the $Am$ and $Re$ active region hatched in white, while the region enclosed by the orange curve is dead according to the $Re$ criterion alone.  In the ``TT Max'' plot on the bottom, it can be seen that nearly the entire bulk mass of the disk is dead based upon $Re$ and $Am$, while the $Re$-only dead region spans just the central $\sim$ 100 AU.  The $Re$-only region can thus be thought of as the minimum size of the dead region, depending upon the importance of $Am$.  We compute the mass contained in the dead zone for each of these scenarios (see Table~\ref{tab:deadmass}).  Recall that the disk mass $M_d = 0.039 M_\odot$, so that the dead zone represents $\sim 1/8$ of the disk (for ISM and solar models) to $\sim 3/4$ of the disk (for the maximum TT model).

\begin{centering}
\begin{deluxetable}{lcc}
\tablecolumns{3} 
\tablewidth{0pt}
\tablecaption{Mass contained in dead zones for the different cosmic ray models without radionuclide ionization, see Fig.~\ref{fig:winddead}.   \label{tab:deadmass}}   
\tabletypesize{\footnotesize}
\tablehead{
  \colhead{Model}  & 
  \colhead{$M_{\rm dead}$: $Re$ \& $Am$}   &
  \colhead{$M_{\rm dead}$: $Re$-only }   \\
  \colhead{}   &              
  \colhead{[$M_{\odot}$]} & 
  \colhead{ [$M_{\odot}$]}                               
}
\startdata
ISM M02 & 0.0038 &     0.0038  \\
Solar Min & 0.0051 &     0.0051 \\
Solar Max & 0.0065 &     0.0065 \\
T-Tauri Min & 0.0268 &     0.0085 \\
T-Tauri Max & 0.0279 &     0.0093 
\enddata
\end{deluxetable}
\end{centering}

The size of the dead zone depends directly on the mass of the disk, which sets the vertical column density of gas normal to the plane of the disk.  A denser (and thus more massive) disk of the same size would thus be less permeable to cosmic rays and have a larger (more massive) dead zone.  Our disk within 400 AU contains 0.039 $M_{\odot}$ of gas and therefore is on the more massive end of the typical range of disk masses.  If we reduce the disk mass by a factor of two, the dead zone shrinks radially by approximately 15\% for the ISM through solar cosmic ray models, but has only a minuscule effect on the extensive dead regions in the T-Tauri models, flattening them in vertical extent by no more than a few AU at $R =$ 100 AU.

In all of the models discussed above, the cosmic ray rate is uniformly reduced throughout the disk.  As discussed in Section~\ref{sec:incidspec}, however, there is an observed radial gradient in the cosmic ray flux, which is both strongly energy and distance dependent.  For example, a $E_{\rm CR} \sim$ 20 MeV cosmic ray at $R =$ 10 AU varies by $\sim$ 0.8\%/AU while the same cosmic ray at  $R =$ 70 AU has an intensity gradient of $\sim$ 9\%/AU \citep{caballerolopez2004}.  This is compared to a $\sim 1$ GeV CR proton, which has uniform gradient of $\sim$ 1\%/AU throughout the Heliosphere.  
To approximate this in our current framework, we have taken the T-Tauri (max)imum extrapolation model and applied a uniformly increasing 2\%/AU radial gradient in the cosmic ray ionization rate, shown in Figure~\ref{fig:winddead}b.  The gradient is incorporated with the expression $\zeta_{\rm CR}(R_{\rm AU}) = \zeta_{\rm CR}$(1 AU)$\times (1+p)^{R_{\rm AU}-1}$, where $\zeta_{\rm CR}$(1 AU) is the wind-modulated cosmic ray ionization rate at $R = 1$ AU (see Section~\ref{sec:zetanh}), $p$ is the fractional cosmic ray increase per AU (thus $p = 0.02$ corresponds to 2\%/AU) and $R_{\rm AU}$ is the distance from the central star in AU. Since the T-Tauri-modulated cosmic ray spectra peak at around $E_{\rm CR} \sim$ 1 GeV where the gradient in the Solar System is closer to 1\%/AU -- 1.5\%/AU, this may over estimate the magnitude of the gradient but nevertheless demonstrates that a modest gradient can allow the outer disk ($R > 250$ AU) to be sufficiently ionized for MRI turbulence even with a strongly modulated cosmic ray intensity.

\subsection{Dead Zone Dependence on the Stellar X-ray Luminosity}\label{sec:extremeX}
\begin{figure}
\begin{centering}
\includegraphics[width=2.5in]{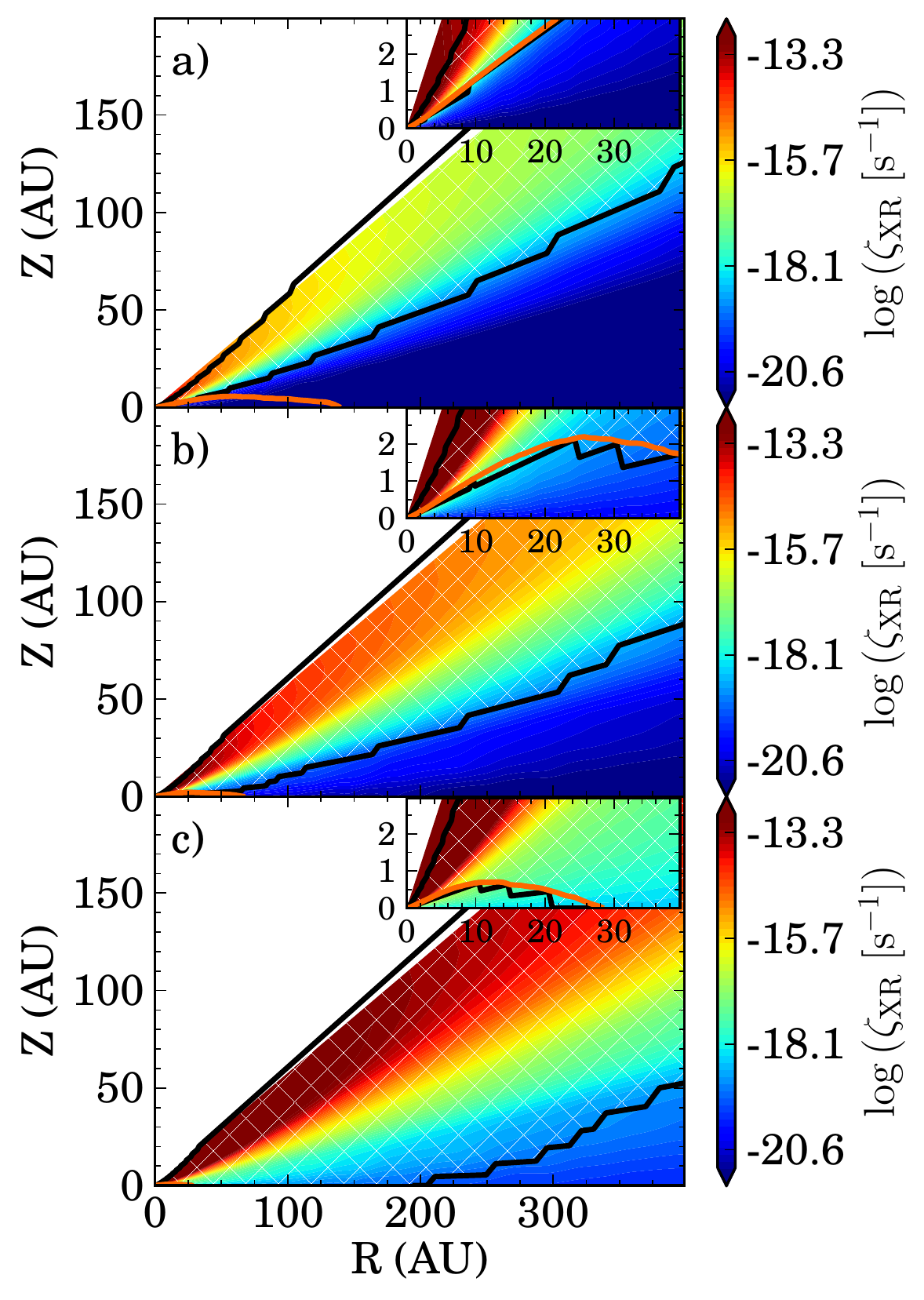}
\caption{Plots of the X-ray only ionization rate throughout the disk per H$_2$, varying the stellar X-ray luminosity: $L_{\rm XR} = $ a) $10^{28}$ erg s$^{-1}$ b) 10$^{29.5}$ erg s$^{-1}$ c) 10$^{31}$ erg s$^{-1}$.  Filled contours show the X-ray ionization rate, $\zeta_{\rm XR}$ s$^{-1}$.  Hatched region and contour lines are the same as for Figure~\ref{fig:winddead}. The right panel, $L_{\rm XR}$ = 10$^{31}$ erg s$^{-1}$, has a harder X-ray spectrum characteristic of flaring T-Tauri stars, see Sections \ref{sec:ionmethod} and \ref{sec:extremeX}. }  \label{fig:xrayL}
\end{centering}
\end{figure}

T-Tauri stars are both X-ray luminous and highly variable \citep{feigelson1981b,feigelson1993,neuhaeuser1995,telleschi2007}. Furthermore, it has been observed that objects with high stellar X-ray luminosity ($L_{\rm XR} \gtrsim 10^{31}$ erg s$^{-1}$) typically have harder X-ray spectra at energies exceeding 2 keV, characteristic of emission contribution from stellar flares \citep{carkner1996,imanishi2001,wolk2005,preibisch2005}. Figure~\ref{fig:xrayL} demonstrates the effect of an enhanced X-ray luminosity on the ionization rates and the resulting size of dead zones. The filled color contours show the ionization rates per H$_2$ for three purely X-ray models (no cosmic rays) with the following X-ray luminosities: $L_{\rm XR}$ = 10$^{28}$, 10$^{29.5}$, and 10$^{31}$ erg s$^{-1}$. In this plot, the crosshatching follows the same convention as Figure~\ref{fig:winddead}. 

For the stellar sources with $L_{\rm XR}$ = 10$^{28}$ and 10$^{29.5}$ erg s$^{-1}$, we adopt the ``characteristic'' spectral template shown in black in Figure~\ref{radfield}, and for the $L_{\rm XR}$ = 10$^{31}$ erg s$^{-1}$ we adopt the ``flaring'' spectral template shown in the same figure in grey to simulate the X-ray hardening with increasing luminosity.  Hard X-ray photons ($E >$ 2 keV) are able to penetrate dense gas more readily than their soft X-ray counterparts, allowing flaring stars to ionize a larger fraction of the disk mass. 

For our prototypical model, $L_{\rm XR}$ = 10$^{29.5}$ erg s$^{-1}$, the disk midplane is largely dead to MRI without inclusion of cosmic ray ionization.  Even in the highly X-ray irradiated case, there are two dead midplane regions, one extending out to $R \sim$ 25 AU (see Fig.~\ref{fig:xrayL} c, inset) and one in the outer disk beyond $R \sim$ 200 AU.  This structure is a result of the $Am$ criterion depending on both density and ionization; in the inner dense disk the ionizing radiation cannot penetrate and in the outer low density disk ion-neutral collisions are not frequent enough. 

This large dead region is contrary to the results of \citet{igea1999} [IG99], who found that incorporation of stellar X-ray scattering into the radiative transfer permitted the entire disk beyond 5 AU to support MRI even in the absence of cosmic rays.  The difference between their results and ours is readily explained by different assumptions for an active disk.  In IG99 the critical electron fraction depends on the viscous disk parameter as $x_{\rm cr} \propto \alpha^{-1/2}$, where $\alpha$ was originally taken to be unity (see Equations 22 and 23 of that work).  Under the same assumptions with $L_{\rm XR}$ = 10$^{29}$ erg s$^{-1}$, our model finds a similarly-sized dead zone, extending out to 5.5 AU in the midplane.  A lower value of $\alpha \simeq $ 0.01 causes MRI to be less efficient at viscously stirring the disk, and therefore the electron fraction required to be turbulent is subsequently higher, creating a substantially larger dead region.

We note that a more dust-settled disk would also change the size of the dead zone by reducing the X-ray opacity due to dust by up to a factor of two \citep{bethell2011a}.  Reducing the X-ray opacity causes deeper layers in the disk to see X-ray photons having a similar effect to increasing the X-ray luminosity.  This would vertically flatten the MRI inactive region but not reduce the radial extent of the dead zone in the midplane.

\section{Further Considerations}
\label{sec:morestuff}

\subsection{Radionuclide Ionization in the Midplane}
\label{sec:radnuc}
An important source of ionization we have neglected thus far is the decay of short-lived radionuclides (RNs).  Species such as $^{26}$Al and $^{60}$Fe have relatively short half-lives, $\tau_{\rm 1/2} < $ 10 Myr, and therefore their presence in the early Solar System is inferred by measurements of the decay products in meteorites \citep[][ and others]{gray1974,macpherson1995}.  While the net ionization rate from RN depends on assumptions regarding their distribution and abundances, typical values estimated for the Solar Nebula from $^{26}$Al decay lie near $\zeta_{\rm RN} = (7.3 - 10) \times10^{-19}$ s$^{-1}$ \citep{umebayashi2009}.  Thus $\zeta_{\rm RN}$ in the early Solar System likely rivals or even exceeds $\zeta_{\rm CR}$ by orders of magnitude for nearly all of the wind-modulation models considered here.  Even the unattenuated solar maximum cosmic ray rate with a uniform 1.5\%/AU CR gradient (Section~\ref{sec:totzone}) has $\zeta_{\rm CR}(R) \lesssim \zeta_{\rm RN}$ within $R \le 100$ AU, assuming the Solar Nebula value of $\zeta_{\rm RN}$ (see Figure~\ref{fig:zetanh} dashed black line). 

\begin{figure*}
\begin{centering}
\includegraphics[width=6.0in]{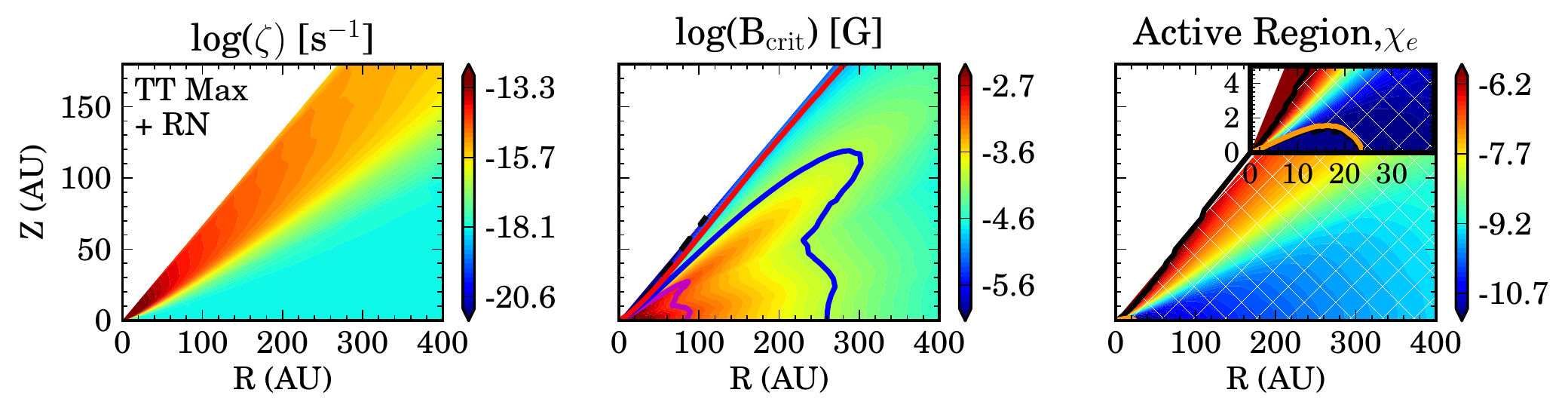}
\caption{Quantities same as Fig.~\ref{fig:winddead}, now including uniform ionization from decay of short-lived radionuclides, $\zeta_{\rm RN} = 7.3 \times 10^{-19}$ s$^{-1}$. The total ionization rate of H$_2$ is given by $\zeta_{\rm tot} = \zeta_{\rm RN} + \zeta_{\rm CR}+\zeta_{\rm XR}$. \label{fig:winddeadnuc}}
\end{centering}
\end{figure*}

Consequently, short-lived radionuclides such as $^{26}$Al may become the dominant source of ionization if they are indeed present in an isolated ($G_0 \sim 1$) protoplanetary disk.  To demonstrate the effect of the addition of ionization by radionuclide decay at levels inferred in the early Solar System, we have recreated the bottom panel (TT Max) of Figure~\ref{fig:winddead} to include contribution from decay of $^{26}$Al by treating it as a uniformly well-mixed ionization source throughout the disk with a magnitude of $\zeta_{\rm RN} = 7.3 \times 10^{-19}$ s$^{-1}$ (Figure~\ref{fig:winddeadnuc}).  In this case the midplane ionization is clearly dominated by ionization as a result of the local radionuclide decay.  For comparison, the presence of radionuclides has little effect on the size of dead zones for the ISM models and moderately restricts the extent of the dead zone to $R \sim$ 20 -- 30 AU for the Solar System-like cosmic ray models.  For the TT Max case, the dead region encompasses the inner $R \lesssim$ 30 AU and contains 0.015 $M_{\odot}$ of material in the dead zone. 

Unfortunately, the ionization contribution from radionuclides in a ``typical'' disk however is entirely unknown. Indeed, there is evidence for an {\it enhanced} abundance of short-lived RN in the early Solar System.  However, the source of this enhancement is exceedingly controversial, typically falling into two categories:  stellar spallation (internal) \citep[e.g.,][]{lee1998,shu2001} or enrichment of the parent molecular cloud from supernovae or Wolf-Rayet stars (external) \citep{wasserburg2006,gounelle2009,gaidos2009,makide2011,gounelle2012}.  While $^{26}$Al can be explained both by internal and external mechanisms, the presence of $^{60}$Fe is solely a stellar nucleosythensis product and thus is external in origin.  Indeed, an inferred enhanced abundance of $^{60}$Fe found in chondrites was originally attributed to a nearby type II supernova during the formation of the Solar System \citep{tachibana2006}; however, this result has come under intense scrutiny with later works claiming no such enhancement \citep{moynier2011,dauphas2011,tang2012}.  In summary, the relative contribution from external versus internal processes is still unclear \citep[for an extensive discussion see ][ and references therein]{adams2010}, but we know that the early Solar System was to some degree enriched by external sources.  In Figure~\ref{fig:zetanh} we include the radionuclide ionization rate as is computed from the mean interstellar abundance of $^{26}$Al derived from $\gamma$-ray observations of $^{26}$Al decay in the Milky Way, \citep[see discussion in][]{umebayashi2009}. This rate is a factor of $\sim$ 8 below the Solar Nebula value, lending credence to the hypothesis that the Solar System formed in an enriched environment.  Furthermore \citet{diehl2006} find the $^{26}$Al abundance to be anisotropic throughout the galaxy, concentrated near massive star-forming regions.  Therefore this enhanced radionuclide abundance may be typical of protoplanetary disks formed in massive clusters.  

\section{Conclusions}
\label{sec:disc}

In this work we have explored several mechanisms that act to reduce the cosmic ray flux incident on protoplanetary disk surfaces.  The first mechanism, exclusion by a magnetized wind, actively operates within the Solar System and excludes cosmic rays at high efficiency, especially at low $E_{\rm CR}$. We have extrapolated the magnitude of cosmic ray exclusion using spot coverage as a tracer of solar magnetic activity to T-Tauri stars with spot coverages of 2\% and 8\%, and have examined the effects on the cosmic ray energy spectrum $J(E)$.  For our extrapolated modulation models we find good agreement with numerical and analytical models of \citet{svensmark2006} and \citet{cohen2012} for their models of the young Sun (Y.S.), where we have extrapolated their results at energies $<$ 10 MeV.  If this reduction in the cosmic ray flux is real, the incident cosmic ray flux should be reduced by at least ten orders of magnitude for 1 MeV cosmic rays and five orders of magnitude at 1 GeV.   In the analysis of \citet{turner2009}, the authors attempt to account for cosmic ray modulation by considering only cosmic rays with energy above 100 MeV, where solar modulation is small.  It can be seen, however, from our results as well as the young Sun models that cosmic ray modulation operates at all energies when wind exclusion is significant (see Fig.~\ref{fig:allcr}).  

Using the results of the cosmic ray propagation models of \citet{padovani2009} we reconstruct the ionization rate versus depth into a molecular slab (disk surface) for the wind-modulated cosmic ray spectra (Fig.~\ref{fig:zetanh}).  We provide numerical fits for the integrated cosmic ray ionization rate $\zeta_{\rm CR}(N_{\rm H_2})$ s$^{-1}$ for a range of modulation strengths.  At the low ionization rates inferred, cosmic rays do not contribute significantly to the ionization rate in the outer disk.  Indeed, the calculated CR rates are many orders of magnitude lower than the ionization rates inferred from decay of short-radionuclides in the early Solar System, though to what degree radionuclides contribute to other systems is unknown.  Regardless, in our simple prediction for wind modulation in a T-Tauri star system we find that the cosmic ray ionization rate is more than an order of magnitude below the interstellar averaged $^{26}$Al ionization rate, $\zeta_{\rm RN} = 9.2 \times 10^{-20}$ s$^{-1}$ \citep{umebayashi2009}.  Therefore, it is this value that we recommend as a minimum H$_2$ ionization rate in disks, though there may be some variation with galactic location \citep[e.g.][]{diehl2006} and with disk evolution, i.e. enhancement via dust settling \citep{umebayashi2013}.

Furthermore, external radiation fields can contribute to the \ion{S}{2} ion abundance in the outer disk, but its contribution is limited by dust extinction if some amount of small dust is present.  Elevated external FUV fields present in stellar clusters, however, can dominate cosmic rays, X-rays and radionuclides at the outermost surface of the disk if the cosmic ray rate is reduced (Fig.~\ref{fig:isrf}).  In the absence of cosmic rays, a strong external field, and radionuclides, the scattered stellar X-rays dominate the H$_2$ ionization in the outer regions of the disk.

In summary, we draw the following conclusions from our models:
\begin{enumerate}[i.]
\item Modulation by stellar winds can reduce the cosmic ray flux in the circumstellar environment by many orders of magnitude, resulting in cosmic ray ionization rates substantially lower ($\zeta_{\rm CR} \lesssim 10^{-20}$ s$^{-1}$) than typically assumed in models of MRI turbulence and circumstellar chemistry.   At the low CR rates inferred, the dominant source of ionization at the midplane throughout the disk can become short-lived radionuclides, if present.
\item If a T-Tauri star drives a Heliosphere-like region of cosmic ray exclusion, the cosmic ray ionization rate should be no higher than $\zeta_{\rm CR} \lesssim 1.4 \times 10^{-18}$ s$^{-1}$, the solar minimum modulation rate. This CR upper limit, however, is likely much too high given the more extreme wind and magnetic properties of T-Tauri stars compared to the Sun, thus we expect $\zeta_{\rm CR} << 10^{-18}$ s$^{-1}$. At the lower end, however, decay by short-lived radioactive particles should provide a floor to the H$_2$ ionization rate set by $^{26}$Al decay, $\zeta_{\rm RN} \gtrsim 9.2 \times 10^{-20}$ s$^{-1}$ at the mean interstellar $^{26}$Al abundance \citep{umebayashi2009}.  As a result, we recommend the inferred $^{26}$Al radionuclide ionization rate in the Solar Nebula for the H$_2$ ionization rate throughout protoplanetary disks, $\zeta_{\rm RN} = (7.3 - 10) \times 10^{-19}$ s$^{-1}$ \citep{umebayashi2009} within a T-Tauriosphere. This is of course in addition to X-ray ionization from the central star. Outside of the T-Tauriosphere the cosmic ray rate should be that of the ISM, and here we suggests ionization rates in the range of the W98 and M02 models (Sections~\ref{sec:crgen} and~\ref{sec:zetanh}).
\item These rates can and will be tested with spatially and spectrally resolved observations of molecular ions with ALMA, at which point more complex models of a T-Tauriospheric cosmic ray modulation will certainly be of interest.  We defer predictions for chemical effects until Paper II.
\item Based on our models we provide fits to the ionization rate $\zeta_{\rm CR}(N_{\rm H_2})$ predicted for T-Tauri stars assuming: 1. extrapolated interstellar CR fluxes (i.e.,  unmodulated), 2. solar wind modulated CR fluxes, and 3. T-Tauri-like wind modulated CR fluxes.  The analytical fits are in the same form as provided by \citet{padovani2009,padovani2013} and are made for ease in use in a variety of physical and chemical simulations.
\item Global magnetic fields can modulate the cosmic ray flux impacting the disk through the competing processes of mirroring and funneling.  The effect is limited to a net 50\% reduction in addition to any wind exclusion but will imprint a 50\% radial gradient in the CR ionization rate.  The possibility of such an effect will need to be considered when modeling ``T-Tauriospheres'' around other stars, as they too will imprint gradients in the cosmic ray flux.
\item  Under these low cosmic ray conditions, large regions of the disk will be unable to sustain MRI turbulence.  Radionuclides can ``reactivate'' MRI turbulence in the midplane outside of $\sim$ 25 AU if they are present at the enhanced rates inferred for the Solar Nebula. Furthermore, an enhanced external FUV radiation field can create an active ``shell'' of material on the disk's outermost surface. 
\end{enumerate}

This paper represents only the first step in assessing the exclusionary effects of winds and magnetic fields in T-Tauri systems. Our work to date utilizes simple, but physically motivated, models. More sophisticated theoretical work should thus be carried out as our understanding of the problem increases. Moreover, the effects explored in this paper will have clear chemical implications that can be readily observed in the near future. As one example, recent studies have sought to measure the degree of turbulence in protoplanetary disks by observing line broadening of strong gas emission lines \citep[e.g.,][]{hughes2011,guilloteau2012}. While we expect vertical stratification between the active surface and dead midplane, it would not be unexpected to find additional radial variations in, for example, turbulent line broadening and coagulative dust growth. In general, protoplanetary disks will be affected by a diversity of ionization sources: stellar UV photons in the inner disk, X-rays in the molecular layer, external radiation ionizing the outer ``skin'' of the disk, and (potentially) both short-lived radionuclides and cosmic rays in the deep planet-forming midplane. Furthermore, all of these ionization sources are likely to vary --- perhaps substantially --- from system to system. As a result, the determination of the diverse ion chemistry of these disks, as well as the turbulent kinematic properties, will be fertile ground for future ALMA observations.

\acknowledgments{The authors are grateful to the referees for their useful suggestions and perspectives.  The authors also wish to thank Mark Moldwin and Marco Fatuzzo for illuminating discussions on various aspects of the paper, M. Padovani for being extremely helpful in providing access to his numerical results, and T. Harries for kindly providing the TORUS radiative transfer code used to produce the disk model.  LIC thanks Prof. Jon M. Miller for assisting with the {\it XSPEC} X-ray modeling. Wilcox Solar Observatory data used in this study was obtained via the web site http://wso.stanford.edu at 2012:11:06, 08:43:00 EST courtesy of J.T. Hoeksema. This work was supported by NSF grant AST-1008800.  FCA is supported by NSF grant DMS-0806756 from the Division of Applied Mathematics and by NASA grant NNX11AK87G from the Origins of the Solar System Program. }

\appendix
\section{Approximation for the Interstellar Radiation Field}
\label{app:isrf}

To determine the extent of the ionization contribution from the ISRF we take a ``brute-force'' approach and coarsely sample our model on a cartesian grid.  From each point in this subsample, we integrate outward along lines of sight evenly spaced over 4$\pi$ steradians assuming azimuthal disk symmetry.  To sample 4$\pi$ steradians uniformly is not trivial, and we adopt here the ``spiral-point technique,'' see \citet{saff1997} for further details and specifically their Equation 8. From these $N_a$ sampled rays we compute a weighted effective optical depth due to extinction by dust,

\begin{equation}
\tau_{\rm eff} = -\ln{\left[ \displaystyle\sum\limits_{a}^{N_{a}} \frac{1}{N_{a}}\exp{(-\tau_a)}\right]}, \label{eq:taueff}
\end{equation} 
in all directions.  The resulting flux is then simply $F_{\rm ISRF} = 1.6 \times 10^{-3}~G_0 \exp{(-\tau_{\rm eff})}$ erg s$^{-1}$ cm$^{-2}$. The expression above yields a scalar measure of the interstellar UV absorption by dust from all directions to a given point in the disk.  Note that this method includes only 1D absorption in a given direction and neglects dust scattering.

\section{Sulfur Chemistry}
\label{app:sulf}
Even though sulfur is less abundant than carbon, it less effectively self-shields and therefore sees the interstellar FUV field deeper in the disk than carbon.  At the temperatures present in the outer disk, $T_g \lesssim 20$ K, most of the S will be locked up in SO and CS on grains and carbon in CO ice.  Visualizing the outer disk as an irradiated slab, there are two dominant ionization fronts: the CO(gr) $\rightarrow$ CO $\rightarrow$ C $\rightarrow$ \ion{C}{2} and CS/SO(gr) $\rightarrow$ CS/SO $\rightarrow$ S $\rightarrow$ \ion{S}{2}.  The location of the ion--neutral transition depends on the intensity of the photodesorbing and photodissociating FUV flux and the recombination rates. For continuum photo-processes, the determining factor is dust attenuation in the outer disk if small grains are present.  
Because CO is very efficient at self shielding, the carbon ionization front traces a thin ``onion-layer'' of \ion{C}{2} near the disk surface.  While sulfur is less abundant than carbon, it does not self-shield as efficiently \citep[in fact, small grains are more efficient, see ][]{pbc2011b}, and as a result, sulfur can be ionized at higher column densities than carbon, setting the deep ionization-front at the UV dust-attenuation limit. Thus, while stellar FUV photoionization of carbon dominates the surface layers and the inner disk, sulfur supplies the FUV-derived ions in the outer disk.

\begin{deluxetable}{lclr}
\centering
\tablecolumns{4} 
\tablecaption{Reduced sulfur network adapted from the OSU gas-phase chemical network \citep[March 2008;][]{smith2004}. \label{tab:sulfnet}}
\tablewidth{0pt}
\tabletypesize{\footnotesize}
\tablehead{
  \colhead{Reactants}  & 
  \colhead{}&
  \colhead{Products}&
  \colhead{Rate $\kappa$ (cm$^{3}$ s$^{-1}$ unless noted)}}
\tabletypesize{\footnotesize}
\startdata
$\mbox{H}_3^+ + \mbox{CO} $ & $ \longrightarrow $ & $ \mbox{HCO}^+ + \mbox{H}_2$ & $\kappa_1 = 1.61 \times 10^{-9}$ \\
$\mbox{H}_3^+ + \mbox{O} $ & $ \longrightarrow $ & $ \mbox{H}_3\mbox{O}^+$ & $\kappa_2 = 8.00 \times 10^{-10}$  \\
$\mbox{S} + \mbox{OH} $ & $ \longrightarrow $ & $ \mbox{SO} + \mbox{H}$ & $\kappa_3 = 6.60 \times 10^{-11}$  \\
$\mbox{H}_3^+ + \mbox{e}^- $ & $ \longrightarrow $ & $ \mbox{H}_2 + \mbox{H}~\mbox{or}~\mbox{H} + \mbox{H} + \mbox{H}$ & $\alpha_1 = 6.70 \times 10^{-8} \left[ \frac{T_g}{300~\rm{K}} \right]^{-0.52}$ \\
$\mbox{H}_3\mbox{O}^+ + \mbox{e}^- $ & $ \longrightarrow $ & $ \mbox{OH} + \mbox{H}_2$ &$\alpha_2 = \frac{2}{3} \times 2.60 \times 10^{-7} \left[ \frac{T_g}{300~\rm{K}} \right]^{-0.5}$ \\
$\mbox{\ion{S}{2}} + \mbox{e}^- $ & $ \longrightarrow $ & $ \mbox{S}$ & $\alpha_3 = 3.90 \times 10^{-12} \left[ \frac{T_g}{300~\rm{K}} \right]^{-0.63}$ \\
$\mbox{SO} + \mbox{gr} $ & $ \longrightarrow $ & $ \mbox{SO(gr)}+ \mbox{gr} $ & $ f_1 = 3.14 \times 10^{-10} \left[ \frac{8 k T_g}{\pi m_{\rm SO}} \right]^{0.5} $ \\
$\mbox{S} + \mbox{gr} $ & $ \longrightarrow $ & $ \mbox{S(gr)}+ \mbox{gr} $ & $ f_2 = 3.14 \times 10^{-10} \left[ \frac{8 k T_g}{\pi m_{\rm S}} \right]^{0.5} $ \\
$\mbox{SO(gr)} + \mbox{Temp} $ & $ \longrightarrow $ & $ \mbox{SO} $ & $ R_1 =  1.32 \times 10^{12} \times \exp{\left[ \frac{-3.34 \times 10^{3}~\rm{K}}{T_d} \right]} $ s$^{-1}$ \\
$\mbox{S(gr)} + \mbox{Temp} $ & $ \longrightarrow $ & $ \mbox{S} $ & $ R_2 = 9.29 \times 10^{11} \times \exp{\left[ \frac{-1.10 \times 10^{3}~\rm{K}}{T_d} \right]} $ s$^{-1}$ \\
$\mbox{SO(gr)} +\gamma_{\rm UV} $ & $ \longrightarrow $ & $ \mbox{SO} $ & $\Gamma_{\rm D 1} = 10^{-3} \sigma_{g}/N_{\rm sites}$ s$^{-1}$ \\
$\mbox{S(gr)} + \gamma_{\rm UV} $ & $ \longrightarrow $ & $ \mbox{S} $ & $\Gamma_{\rm D 2} = 10^{-3}  \sigma_{g}/N_{\rm sites}$ s$^{-1}$ \\
$\mbox{SO} + \gamma_{\rm UV} $ & $ \longrightarrow $ & $ \mbox{S}+ \mbox{O}$ & $\Gamma_{\rm SO} = 3.30 \times 10^{-10} G_0$ s$^{-1}$\\
$\mbox{OH} + \gamma_{\rm UV} $ & $ \longrightarrow $ & $ \mbox{O}+ \mbox{H}$ & $\Gamma_{\rm OH} = 1.68 \times 10^{-10} G_0$ s$^{-1}$\\
$\mbox{S} + \gamma_{\rm UV} $ & $ \longrightarrow $ & $ \mbox{\ion{S}{2}}+ \mbox{e}^-$ & $\Gamma_{\rm PI} = 7.20\times 10^{-10} G_0$ s$^{-1}$ 
\enddata
\end{deluxetable}

Carbon is ionized by FUV photons with 912--1109 \AA, while sulfur is ionized for photons of $\lambda$ 912--1198 \AA, which is equivalent to 7\% and 10\% of the interstellar FUV flux between 912 -- 2000 \AA, respectively.  Combining the stellar and interstellar UV fields, we then solve for the steady-state \ion{S}{2} abundance taking into account the pathways and rates listed in Table~\ref{tab:sulfnet} \citep[from the OSU gas-grain network;][]{smith2004}, as well as the total sulfur abundance: $\mbox{S}_{\rm tot} = \mbox{S} +  \mbox{\ion{S}{2}} + \mbox{SO} + \mbox{SO(gr)} + \mbox{S(gr)}$. To simplify the problem the electron abundance is computed from Eq.~(\ref{eq:xe}), which intrinsically assumes most electrons come from ionization of H$_2$, and we furthermore set the oxygen abundance to be $\chi_{\rm O} = 10^{-10}$, consistent with the outer disk where most oxygen is in molecular gas or ice \citep{fogel2011}. We tested values ranging between $\chi_{\rm O} = 10^{-9}-10^{-12}$ and find that the result does not depend strongly on the assumed oxygen abundance. Additionally, we set the CO abundance to be $\chi_{\rm CO} = \chi_{\rm C} \equiv 1.4 \times 10^{-4}$.  In the following equations, number density (cm$^{-3}$) of a particular species is denoted by brackets around the species name, e.g., [CO]. 
First, we compute directly the steady state abundances of H$_3^+$ and H$_3$O$^+$:
\begin{gather*}
 \left[ \rm{H_3^+} \right] = \frac{\zeta_{\rm H_2} [\rm{H_2}]}{\kappa_1 [\rm{CO}]+\alpha_1 [\rm{e^-}]+\kappa_2 [\rm{O}]},\\
 \left[ \rm{H_3O^+} \right] = \frac{\kappa_2 [\rm{H_3^+}][\rm{O}]}{\alpha_2 [\rm{e^-}]}.
\end{gather*}
In steady state, the \ion{S}{2} number density can be rearranged into a quadratic form, A [\ion{S}{2}]$^2$ + B [\ion{S}{2}] + C = 0, where the coefficients are defined as follows: 
\begin{gather*}
\mbox{A} = -\frac{\kappa_3 \alpha_3 [\rm{e^-}]}{\Gamma_{\rm PI}} \left(1+ \mbox{E} \frac{\alpha_3 [\rm{e^-}]}{\Gamma_{\rm PI}} \right), \\
\mbox{B} = \left[ \frac{ \kappa_3 \alpha_3 [\rm{e^-}][S_{\rm tot}]}{\Gamma_{\rm PI}} - \left(1+\mbox{E} \frac{\alpha_3 [\rm{e^-}] }{\Gamma_{\rm PI}} \right) \Gamma_{\rm OH} -   \mbox{D} \frac{\alpha_2 \alpha_3 \kappa_3   \left[ \rm{H_3O^+} \right] [\rm{e^-}]^2}{\Gamma_{\rm PI} \Gamma_{\rm SO}} \right], ~\rm{and} \\
\mbox{C} = \Gamma_{\rm OH} [\rm{S_{tot}}]; 
\end{gather*}
with D and E defined as:
\begin{gather*}
\mbox{D} = \left[1+ \frac{ f_1 [\rm{gr}]}{\Gamma_{\rm D 1} + R_1} \right]~\rm{and} \\ 
\mbox{E} = \left[1+ \frac{ f_2 [\rm{gr}]}{\Gamma_{\rm D 2} + R_2} \right].  
\end{gather*}
Upon solving the quadratic equation for [\ion{S}{2}], the corresponding abundances of the other species can be computed straightforwardly:
\begin{gather*}
\left[\mbox{S}\right] = \frac{\alpha_3 [\mbox{\ion{S}{2}}][\rm{e^-}]}{\Gamma_{\rm PI}},\\
\left[\mbox{OH}\right] = \frac{\alpha_2 [\rm{H_3O^+}][\rm{e^-}]}{\kappa_3 [\rm{S}] + \Gamma_{\rm OH}}, \\
\left[\mbox{SO}\right] = \frac{\kappa_3 [\rm{OH}] [\rm{S}]}{\Gamma_{\rm SO}}, \\
\left[\mbox{SO(gr)}\right] = \frac{f_1 [\rm{SO}] [\rm{gr}]}{\Gamma_{\rm D 1} +R_1},~\rm{and}\\
\left[\mbox{S(gr)}\right] =  \frac{f_2 [\rm{S}] [\rm{gr}]}{\Gamma_{\rm D 2} +R_2}. 
\end{gather*}

In this treatment, we have significantly simplified the sulfur chemistry by excluding the formation of other S-bearing species such as CS.  In general, by allowing sulfur to take other pathways to form species such as CS, more sulfur will be tied up in gas or ice phase molecules and will be less available for photoionization, pushing the ion-front closer to the disk surface. For a detailed treatment of this problem, both the chemistry and multidirectional self-shielding of molecules such as CO and sulfur need to be considered in detail.  This approximation, however, gives a simple expression for the ionization contribution from \ion{S}{2} in the outer irradiated region in the disk.
  
\end{document}